%% file: main.tex
\documentclass[sigconf, nonacm, pdfa]{acmart}
\usepackage[a-2b]{pdfx}
\usepackage{soul}
\usepackage{xspace}
\usepackage{multirow}
\usepackage{booktabs}
\usepackage{siunitx}
\usepackage[dvipsnames,table]{xcolor}
\usepackage[utf8]{inputenc}
\usepackage[T1]{fontenc}          
\usepackage[scaled=0.85]{beramono}
\usepackage{listings}
\usepackage{booktabs}
\usepackage{mdframed}
\usepackage{makecell}
\usepackage{enumitem}

\definecolor{lightyellow}{RGB}{255,255,204}
\definecolor{highlight}{RGB}{255,255,100}

\setlist[itemize]{leftmargin=1em, labelsep=0.3em, nosep}
\setlist[enumerate]{leftmargin=1.5em, labelsep=0.3em, nosep}

\sethlcolor{highlight}
\lstdefinestyle{sqlpretty}{
  language=SQL,
  backgroundcolor=\color{lightyellow},
  escapeinside={(*@}{@*)},
  moredelim=[is][\hl]{(*@}{@*)},
  morekeywords={WITH,AS,SELECT,DISTINCT,FROM,JOIN,LEFT,RIGHT,INNER,OUTER,FULL,
                ON,WHERE,AND,OR,NOT,IN,IS,NULL,BETWEEN,LIKE,EXISTS,
                GROUP,BY,HAVING,ORDER,LIMIT,OFFSET,UNION,INTERSECT,EXCEPT,
                 CREATE,TEMP,TEMPORARY,TABLE,VIEW,INSERT,INTO,VALUES,UPDATE,
                SET,DELETE,CASE,WHEN,THEN,ELSE,END,CAST,COALESCE},
  keywordstyle=\color{blue}\bfseries,
  basicstyle=\ttfamily\small,
  breaklines=true,
  breakatwhitespace=false,
  columns=fullflexible,
  keepspaces=true,
  captionpos=b,
  literate={timestamp}{{\textcolor{black}{\textnormal{\ttfamily timestamp}}}}{9},
}

\graphicspath{{figures/}}

\newcommand{\sysname}{MetaSieve}
\newcommand{\hide}[1]{}

\newcommand{\besttime}[1]{\cellcolor{green!20}#1}
\newcommand{\meta}[1]{$\langle$\textsf{#1}$\rangle$}

\begin{document}
\title{MetaSieve: Faster Relational Deep Learning through SQL-Based Metapath Selection}

\author{Fahim Shahriar Khan}
\affiliation{%
  \institution{University of Texas at Arlington}
}
\email{fsk2739@mavs.uta.edu}

\author{Ashraf Aboulnaga}
\affiliation{%
  \institution{University of Texas at Arlington}
}
\email{ashraf.aboulnaga@uta.edu}

\begin{abstract}

\input{\secdir abstract}
\end{abstract}

\maketitle

\pagestyle{plain}

\vspace{-0.80\dbltextfloatsep}
\input{\secdir intro}

\input{\secdir related}
\input{\secdir probdef}
\input{\secdir sql}

\input{\secdir scoring}
\input{\secdir experiments}

\input{\secdir conc}

\bibliographystyle{ACM-Reference-Format}
\bibliography{ref}

\end{document}

%% file: sections/abstract.tex
Relational Deep Learning (RDL) is an effective approach to machine learning over multi-table relational databases. In RDL, a database is modeled as a graph in which each row is a node and each foreign-key relation is an edge, and a graph neural network (GNN) is trained on this graph. Training a GNN requires sampling a subgraph around every seed node in the training set, and the cost of training is largely determined by the size of these subgraphs. This paper aims to reduce subgraph size by leveraging the join and aggregation capabilities of relational database systems. We observe that sampled subgraphs are obtained by following metapaths composed of foreign-key links, and that many of these metapaths can be pruned without loss of accuracy. We present \sysname{}, a metapath selection layer that determines which metapaths to retain and which to prune. For each candidate metapath extension, \sysname{} computes statistics via SQL join and aggregation queries and evaluates the extension based on a novel scoring function that prefers lightweight but informative candidates. Metapaths whose scores fall below a threshold are deemed uninformative and pruned. Metapath selection in \sysname{} is lightweight since it relies only on database statistics and task labels, and it is independent of GNN parameters, so it integrates with diverse GNN architectures for classification and regression. Our evaluation on the RelBench benchmark with multiple GNN backbones shows that \sysname{} consistently reduces per-epoch training time by large margins while maintaining and often improving accuracy.

\vspace{4pt}
\noindent
\textbf{Code Availability:}\\
\url{https://github.com/lids-lab/metasieve}

%% file: sections/intro.tex
\vspace{10pt}
\section{Inroduction}
\label{sec:intro}

Relational databases remain the default foundation for operational and analytical data, and much of the information used to train machine learning models is ultimately stored in relational database tables, connected by foreign-key relationships. Relational Deep Learning (RDL) has emerged as a practical way to learn directly from this multi-table structure by modeling the database as a heterogeneous graph and training a graph neural network (GNN) on this graph~\cite{relbenchposition}. Each row in the database becomes a node in the graph whose type corresponds to its source table, and each foreign-key link becomes an edge whose type corresponds to its foreign-key type.
By operating on the native relational structure, RDL sidesteps the need to flatten tables or hand-engineer features, and has shown strong results across multiple prediction tasks~\cite{relbench}.

To train an RDL model for a specific task, the training data consists of rows from some table, termed \textit{seed rows}, with a task-specific label associated with each row.
For example, if the task is predicting customer churn, the seed rows are customers (identified by their customer id) and the labels indicate whether the customer churned. These rows become the \textit{seed nodes} in the RDL graph, and standard GNN architectures are trained by \textit{sampling a subgraph} from the multi-hop neighborhood of each seed node and using the sampled subgraphs in standard backpropagation neural network training~\cite{graphsage, hgt, relgt}.
In the churn example, the sampled subgraph for each customer summarizes their behavior (e.g., purchases, reviews) and is used to train the GNN churn model.

The cost of GNN training depends heavily on the size and complexity of these sampled subgraphs. These subgraphs are typically sampled by randomly selecting neighbors at each hop, which in the RDL context amounts to following random foreign-key links.
One way to improve sampling is to consider the \textit{metapaths} being sampled.
A metapath is a sequence of foreign-key edge types traversed from a seed node, and it is well established in heterogeneous graph learning that not all metapaths are equally useful~\cite{graphmse, pmhgnn,mpgnn}.

Some metapaths are irrelevant to the task, others occur too rarely to provide a useful training signal, and still others are very frequent but noisy, increasing training cost without adding label-relevant information.
An effective way to reduce the cost of GNN training is therefore to identify useful metapaths and restrict subgraph sampling to these metapaths, a problem known as \textit{metapath selection}.

Metapath selection is particularly important for RDL because relational database schemas can be complex, inducing many diverse metapaths of varying usefulness, while large database instances increase the cost of sampling~\cite{mpsgnn,labor}.
In this paper, we introduce \textit{\sysname{} (Metapath Selection via SQL-Informed Evaluation of Valuable Extensions)}, a metapath selection technique for RDL that relies on two key ideas: (1)~collecting metapath statistics using SQL queries, thereby leveraging a strength of the database system~\cite{sqlreldl}, and (2)~identifying uninformative metapaths to prune (``sieve'') using principled scoring based on the statistics and task labels.

Prior work has shown that counting metapath occurrences is important for metapath selection, since the task label is often determined by aggregate evidence obtained from multiple occurrences of a metapath~\cite{mpsgnn}.
\sysname{} counts metapath occurrences using SQL queries that are simple, efficient, and do not require materializing the graph.
For example, consider the \texttt{rel-stack} database from the RelBench benchmark~\cite{relbench}, which represents the online activity of users on Stack Overflow, and consider the metapath \texttt{user$\rightarrow$posts$\rightarrow$comments}.
We can compute the number of occurrences of this metapath for every seed user using a SQL aggregation query (e.g., \texttt{COUNT(comments)}).
We can also compute more complex statistics such as the ratio of comments per post, which we term the \textit{rate}, capturing ``evidence per opportunity'' (e.g., \texttt{COUNT(comments) / COUNT(posts)}).
The queries used by \sysname{} are simple join-and-aggregation queries that can be executed efficiently, and we structure the computation so that each query is executed only once.

After collecting statistics for all candidate metapaths, \sysname{} treats metapath selection as a dependence-ranking problem and uses the mutual information (MI) between the statistics and the task label as a measure of metapath relevance~\cite{peng2005feature}, building on the standard view that MI captures general (including non-linear) dependence and is effective for lightweight feature scoring~\cite{battiti1994mi}.
To penalize candidates that inflate the sampled neighborhood, \sysname{} makes its relevance measure cost-aware by favoring candidates that achieve high MI with low fanout. \sysname{} also penalizes metapaths infrequently observed in the data, and it combines these measures in a principled way to identify uninformative paths that can be avoided while sampling. 
Since MI applies to both discrete and continuous labels, the \sysname{} pipeline supports classification and regression tasks.
Furthermore, the pipeline is agnostic to the GNN architecture, so it can be used with any GNN backbone.

We conduct experiments with multiple datasets from the RelBench benchmark~\cite{relbench}, including the largest, recently added datasets, and we show that \sysname{} substantially outperforms the state-of-the-art MPS-GNN technique~\cite{mpsgnn} in terms of preprocessing time, and \textbf{speeds up training by up to an order of magnitude compared to random sampling and MPS-GNN}. \sysname{} \textbf{often leads to an improvement in accuracy} in addition to reducing training time, especially with the state-of-the-art Relational Graph Transformer (RelGT)~\cite{relgt} GNN. These results hold for both classification and regression tasks.

In summary, our contributions are as follows:
\begin{itemize}
  \item We introduce a task-specific metapath selection pipeline for RDL that computes metapath statistics efficiently using SQL queries, thereby avoiding graph materialization while directly leveraging relational structure.
  \item We propose a novel scoring function for pruning uninformative metapaths that uses mutual information as a lightweight proxy for relevance, penalizes large fanout to favor more compact subgraphs, and downweights infrequently observed metapaths.
  \item We evaluate the full pipeline on both node classification and node regression tasks from RelBench and show that the resulting metapath selection rules transfer across a wide range of GNN architectures, demonstrating the broad applicability of \sysname{}.
\end{itemize}

%% file: sections/related.tex
\vspace*{-6pt}
\section{Related Work}
\label{sec:related}
\noindent
\textbf{Metapath selection in heterogeneous graphs.}
Metapath selection methods for heterogeneous graphs broadly fall into two groups: methods that learn metapath importance jointly with representation learning inside the model, and methods that treat metapath selection as an external search or optimization problem. In the first group, HAN \cite{han} assumes a candidate set of metapaths and learns their importance through hierarchical attention, GTN \cite{gtn} removes the need for manually specified metapaths by learning soft edge-type compositions across hops, and GraphMSE \cite{graphmse} combines efficient metapath sampling with attention-based weighting and semantic feature-space alignment. In the second group, PM-HGNN \cite{pmhgnn} formulates metapath selection as a reinforcement-learning problem over node-specific metapaths, while LMSPS \cite{lmsps} treats metapath discovery as an explicit search problem with progressive sampling. \sysname{} is more like the second group since it treats metapath selection as a preprocessing step before GNN training.

\vspace{4pt}
\noindent
\textbf{Relational deep learning.}
Fey et al.~\cite{relbenchposition} introduce RDL as an end-to-end learning framework for relational databases. 4DBInfer~\cite{wang20244dbinfer} similarly frames predictive modeling on relational databases as graph-centric multi-table learning and emphasizes the joint role of graph construction, subgraph extraction, and model design in a unified toolbox. RelBench~\cite{relbench} provides a public benchmark and the first broad empirical study of RDL.  RelBench v2~\cite{relbenchv2} expands the benchmark by adding larger, more diverse databases, new prediction tasks, and support for pretraining and transfer. Within this emerging research area, several backbone architectures have been proposed. RelGNN~\cite{relgnn} is an early RDL-specific model that addresses message passing challenges across multiple related tables through composite message passing. RelGT~\cite{relgt} is a recent model that uses a transformer-based design over sampled relational neighborhoods and shows remarkable accuracy. Relational Transformers~\cite{relationaltransformer} explore pretraining for transfer to unseen datasets and tasks, extending RDL toward relational foundation models~\cite{kumorfm,griffin}. \sysname{} works with any GNN-based RDL model, and we evaluate it on the state-of-the-art RelGT, in addition to older GNN models.

\vspace{4pt}
\noindent
\textbf{Metapath selection for relational deep learning.}
Two closely related methods for metapath selection in the node classification setting in RDL have recently been proposed.
MP-GNN~\cite{mpgnn} learns useful metapaths incrementally by scoring each extension of a partial metapath, assuming that node labels can be explained by whether a metapath exists in the data.
MPS-GNN~\cite{mpsgnn} shows that this assumption can be overly restrictive, since predictions may depend on aggregate evidence from multiple occurrences of the same metapath. MPS-GNN addresses this limitation by learning a scoring function that implicitly uses occurrence-based information rather than path existence alone. This allows the search to favor metapaths whose occurrence patterns, including higher-order statistical patterns like counts-of-counts, is predictive of the node label. However, the search is costly because assessing candidate extensions requires repeatedly training a GNN on the metapaths selected so far. In our experiments, we found that MPS-GNN does not scale because of the need to train a GNN at each step. We also found that it selects metapaths inferior to \sysname{}. Furthermore, it only works for classification, not regression.

%% file: sections/probdef.tex
\vspace*{-4pt}
\section{Problem Definition}
\label{sec:probdef}

\begin{figure*}[t]
    \centering
    \vspace*{-4pt}
    \includegraphics[width=.75\textwidth]{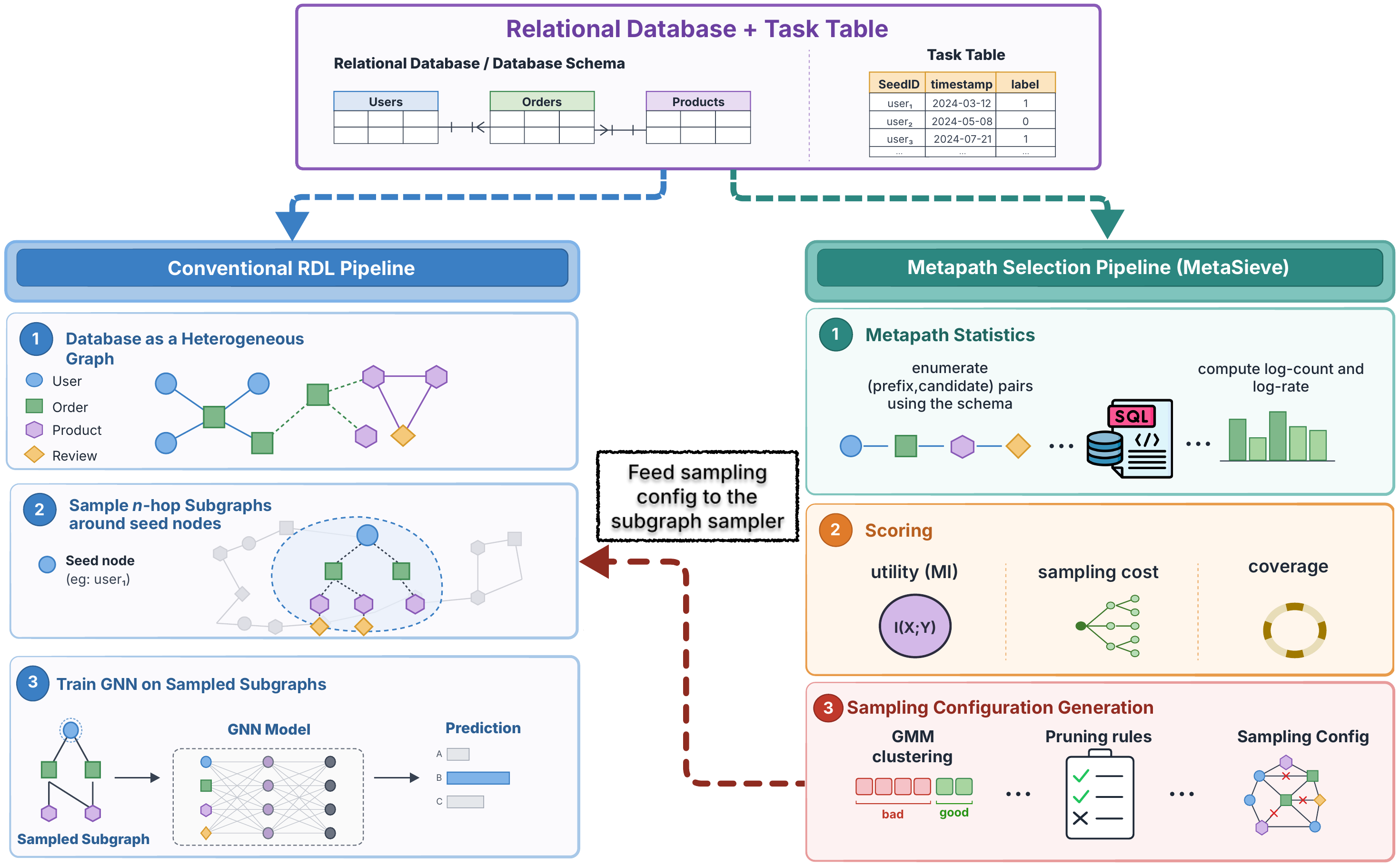}
    \caption{The \sysname{} pipeline. \sysname{} works as a metapath selection layer that plugs into the conventional RDL pipeline. \sysname{} provides sampling configurations to the subgraph sampler, which provides subgraphs for downstream GNN training.}
    \label{fig:pipeline_overview}
    \vspace*{-4pt}
\end{figure*}

We consider predictive tasks defined by a \emph{task table} over a relational database. Each row of the task table defines one training instance, or \emph{seed node} $(s,t,y)$, consisting of an identifier $s$, a timestamp $t$, and a task label $y$. The identifier $s$ is a foreign key linking the task table to the database table that contains the seed nodes, the \emph{seed node source table}. Sampling begins from this table and expands outward along the graph up to a maximum hop limit $H$. To prevent temporal leakage, only traversed rows with timestamp $t' < t$ are sampled. This temporal consistency condition is also used in RelBench. Tables without timestamps require no temporal condition, and all rows in such tables connecting to $s$ are treated as available for sampling. \sysname{} assumes that the database contents, schema, and task definition remain fixed during preprocessing, training, and inference; any changes require rerunning the pipeline. This is a common assumption in RDL, and we leave dynamically handling database or schema updates as future work.

A \emph{metapath} is a sequence of schema-valid foreign-key traversals starting from the seed node source table. We write a metapath of length $h$ as $\pi_{1:h}=(r_1,\ldots,r_h)$, where $h \leq H$ and each $r_i$ is one traversal in the relational graph. As an example, recall the metapath
\texttt{user$\rightarrow$posts$\rightarrow$comments}
from the introduction.

If table $A$ has a foreign key referencing table $B$, the relational graph has a directed edge from the $A$ node to the $B$ node, which we call the \emph{forward edge}. We add a \emph{reverse edge} for every forward edge, connecting the primary-key node $B$ to the foreign-key node $A$.
We call a join that follows a forward edge a \emph{forward join} and a join that follows a reverse edge a \emph{reverse join}. Thus, every metapath $\pi_{1:h}$ can be viewed as a sequence of forward and reverse joins.

For a metapath of length $h$ $\pi_{1:h}=(r_1,\ldots,r_h)$, let $T_h$ denote the \emph{terminal table} of this metapath, i.e., the final table in the join sequence induced by $\pi_{1:h}$. When extending this metapath to length $h+1$, 
we call $\pi_{1:h}$ the \emph{prefix metapath}, and  we define the set of valid next edge traversals from $T_h$ as $\mathcal{C}(\pi_{1:h})$. Each $e \in \mathcal{C}(\pi_{1:h})$ is called a \emph{candidate extension}. Thus, each pair $(\pi_{1:h}, e)$ represents one possible candidate extension of the prefix metapath $\pi_{1:h}$ that \sysname{} evaluates to decide whether it should be pruned or sampled.

\vspace{4pt}
\noindent
\textbf{Problem definition.} Given a maximum hop count $H$, a database schema, and a set of labeled training seed nodes, the goal of \sysname{} is to produce a compact set of sampling rules $\mathcal{R}$. Each rule is defined for one candidate extension, specified by a hop $h\in\{0,\ldots,H-1\}$, a metapath prefix $\pi_{1:h}$, and the candidate extension itself $e\in\mathcal{C}(\pi_{1:h})$.
Each rule specifies whether that extension is expanded or pruned:
$\mathcal{R}(\pi_{1:h},e)\in\{\texttt{expand},\texttt{prune}\}$.

\vspace*{-4pt}
\subsection{Metapath Informativeness}
\label{sec:metapath_informativenss}

Following a metapath from a particular seed node reaches a set of database rows. Although the traversal sequence is fixed by the schema, the number of distinct rows reached can vary substantially across seed nodes. This variation forms the metapath's occurrence pattern and can provide useful predictive evidence when it varies systematically across seed nodes with different task labels. Many relational task labels are influenced by such activity patterns, including the number of purchases, reviews, visits, or comments associated with an entity.

A schema-valid metapath may not provide predictive signal for a particular task. Schema validity only means that the traversal is permitted by the database schema, not that it helps predict the task label. For example, in a user-churn task, a metapath to an account-type table may reach exactly one row for every user, regardless of whether the user churns. Because its occurrence pattern is nearly identical across the two label groups, the metapath provides little discriminative signal. A schema-valid metapath may also occur for too few seed nodes to provide reliable evidence or may introduce many neighbors while contributing little label-relevant information.

MetaSieve therefore considers a metapath useful when its occurrence pattern is discriminative of the task label, with this dependence quantified using mutual information (Section~\ref{sec:mi_proxy}), also when it is observed for enough seed nodes and provides sufficient predictive value to justify its additional sampling cost.

\vspace{-8pt}
\section{Method Overview}
\label{sec:overview}
The \sysname{} pipeline (Figure~\ref{fig:pipeline_overview}) works as follows:
\begin{enumerate}
    \item \textbf{Schema processing.} Scan the schema and enumerate all valid metapaths starting from the seed node source table, up to a maximum of $H$ hops. At every enumeration step, the metapath enumerated thus far is the prefix metapath, and the next step is the candidate extension (Section~\ref{sec:enumeration}).
    
    \item \textbf{SQL.} Compute metapath statistics for each (prefix, candidate extension) pair over the training seed nodes using SQL queries and without materializing the full graph
    (Section~\ref{sec:sql}). The seed nodes are divided into \textit{batches} for this computation (Section~\ref{sec:batchwise_protocol}).

    \item \textbf{Candidate extension scoring.} Combine these metapath statistics with the seed-node labels and metapath frequency to score each candidate extension. The score reflects the extension's task relevance, sampling cost, and support in the data. Extensions with low scores are pruned (Section~\ref{sec:scoring}).
    
    \item \textbf{Sampling rules.} Use the scores to derive pruning rules that specify which candidate extensions should not be sampled. These rules are provided to the sampler and result in a lightweight sampled subgraph for each seed node that still retains the task-relevant evidence needed for GNN training (Section~\ref{sec:rule_generation}).
\end{enumerate}

%% file: sections/sql.tex
\vspace*{-4pt}
\section{Schema-guided Metapath Enumeration}
\label{sec:enumeration}

\sysname{} starts by enumerating all possible metapaths originating from the seed node source table by traversing all schema-valid foreign-key links, forward or reverse, up to a maximum hop count $H$. At hop $h$, for a metapath prefix $\pi_{1:h}$ representing a valid sequence of relation traversals from the seed node source table to terminal table $T_h$, the schema specifies all valid next relation traversals $\mathcal{C}(\pi_{1:h})$. 
For every $e \in \mathcal{C}(\pi_{1:h})$, \sysname{} evaluates the pair $(\pi_{1:h}, e)$.

During enumeration, we disallow immediate backtracking in which a reverse edge is followed by its corresponding forward edge, since this returns to the same node. Such a loop would create candidate extensions that add no new information. However, we do allow metapaths that return to the same node via longer cycles through multiple tables.

\section{Metapath Statistics using SQL}
\label{sec:sql}

\sysname{} collects statistics about each candidate extension $(\pi_{1:h},e)$ from the database tables using SQL queries.
The key idea is to materialize a \emph{frontier table} for each metapath prefix $\pi_{1:h}$ representing all the rows reachable by this prefix, and then to reduce the frontier table to a small set of per-seed-node statistics,
without materializing the relational graph. Each frontier table is computed once and reused for all candidate extensions that have $\pi_{1:h}$ as a prefix.

\vspace*{-4pt}
\subsection{Computing the Frontier}
\label{sec:frontier_representation}

For a labeled training seed node $(s,t,y)$ and a metapath prefix $\pi_{1:h}=(r_1,\ldots,r_h)$ whose terminal table is $T_h$, we define the hop-$h$ frontier $F_h(s,t,\pi_{1:h})$ as the set of distinct row identifiers in $T_h$ that are reachable from $s$ by following the join sequence induced by $\pi_{1:h}$ under the temporal constraint at time $t$.
The frontier retains each reachable row identifier only once. As discussed below, duplicate elimination is required for forward joins, where multiple parent-frontier rows may reach the same terminal row, but is unnecessary for reverse joins.

The frontier $F_h(s,t,\pi_{1:h})$ is materialized as a frontier table. Every frontier table contains three \emph{core} columns: \texttt{SeedId}, \texttt{timestamp}, and \texttt{last\_id}. The pair \texttt{(SeedId, timestamp)} identifies the seed node, and \texttt{last\_id} stores the identifier of a distinct row currently reached in the terminal table $T_h$. Thus, for each seed node, the frontier table contains one row for each row of $T_h$ it reaches, and \texttt{(SeedId, timestamp, last\_id)} forms its primary key. Candidate extensions at the next hop may require following foreign-key links from this row of $T_h$, \emph{so these links are also included in the frontier table}.
That is, the frontier table includes optional columns \texttt{foreign\_key\_1}, \texttt{foreign\_key\_2}, \ldots, \texttt{foreign\_key\_n}. These store the foreign-key values of the current row in the terminal table $T_h$, which will be needed for forward extension at the next hop.

This frontier table schema supports both forward and reverse joins. If the candidate extension is a reverse join, the next table is reached by matching one of its foreign keys to the current \texttt{last\_id}, so no additional information is needed in the frontier table beyond the core columns. If the candidate extension is a forward join, the next table is reached by following a foreign-key value in the current row of $T_h$ to the primary key of a row in the referenced table, and all the necessary foreign-key values are in the frontier table.

The two join directions also differ in whether duplicate elimination is required. In a reverse join, each matching row in the next table
contributes its own unique row identifier, so the result is already
duplicate-free. In a forward join, multiple rows in the parent
frontier may contain the same foreign-key value and therefore reach
the same row in the next table. \sysname{} consequently applies
\texttt{DISTINCT} only to forward joins and not for reverse joins.

\begin{figure}[t]
    \centering
    \includegraphics[width=\columnwidth]{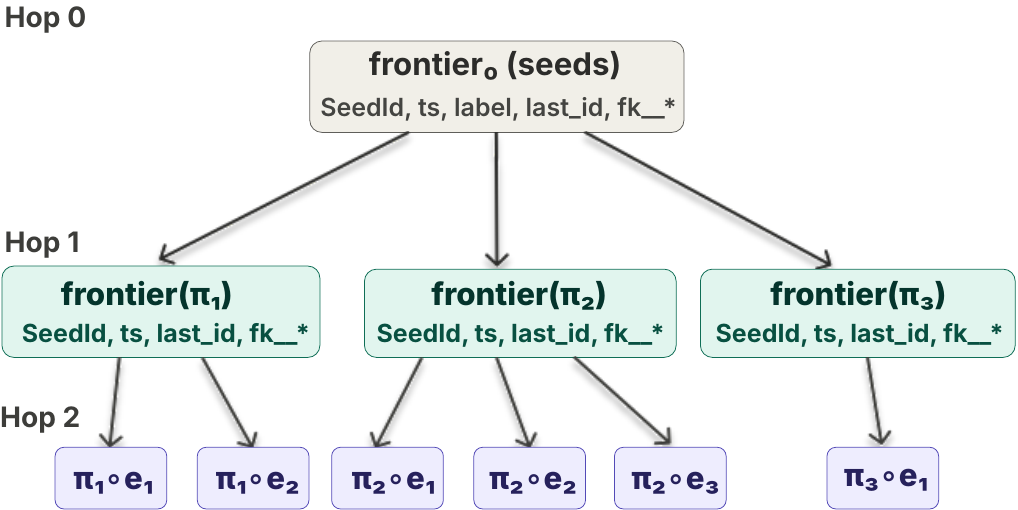}
    \caption{Generating frontier tables across hops. The hop-0 frontier is initialized from the training seed nodes, and each later frontier is generated from its parent by one additional join. Thus, each child hop-$h$ frontier reuses the already-materialized parent hop-$(h-1)$ frontier. $ts$ stands for timestamp.}
    \label{fig:frontier_expansion}
    \vspace*{-8pt}
\end{figure}

We initialize the frontier table using the training seed nodes at hop 0. The exact form of this initial frontier depends on the types of candidate extensions at hop 1. If the hop-1 candidate set contains forward joins, the initial frontier must store not only the seed identifier as \texttt{last\_id}, but also the foreign-key values needed for these forward joins. In this case, we build the hop-0 frontier by a join that attaches each training seed node to its row in the seed node source table and copies the required foreign-key columns (Listing~\ref{lst:seeds-frontier-forward}).

As described in Section \ref{sec:probdef}, \meta{temporal\_constraint} in the WHERE clause is instantiated as $t' < t$ when the joined table has a timestamp $t'$; otherwise, the predicate is omitted. The same rule is applied whenever a timestamped table is joined in subsequent frontier-construction queries, so that frontier rows never contain events that had not yet occurred at the timestamp of the seed node.

If the hop-1 candidate set contains only reverse joins, then no foreign-key columns need to be included in the frontier table. In this case, the initial frontier can store only the training seed node, its timestamp, and the seed identifier as \texttt{last\_id} (Listing~\ref{lst:seeds-frontier-reverse}).

\begin{lstlisting}[style=sqlpretty, caption={Hop-0 frontier when hop-1 includes forward joins.}, label={lst:seeds-frontier-forward}]
CREATE TEMP TABLE frontier_0 AS
SELECT 
    SeedId,
    timestamp,
    SeedId AS last_id,
    (*@\meta{required\_foreign\_key\_1}@*),
    (*@$\ldots$@*),
    (*@\meta{required\_foreign\_key\_m}@*)
FROM training_seed_nodes
JOIN seed_node_source_table
  ON seed_node_source_table.primary_key = training_seed_nodes.SeedId
WHERE (*@\meta{temporal\_constraint}@*);
\end{lstlisting}

\begin{lstlisting}[style=sqlpretty, caption={Hop-0 frontier when hop-1 has only reverse joins.}, label={lst:seeds-frontier-reverse}]
CREATE TEMP TABLE frontier_0 AS
SELECT
    SeedId,
    timestamp,
    SeedId AS last_id
FROM training_seed_nodes;
\end{lstlisting}

After the hop-0 frontier table is initialized, all later frontier tables are generated in the same way, illustrated in Figure~\ref{fig:frontier_expansion}. Each frontier is obtained from an already-materialized parent frontier via one additional join. Thus, for a candidate extension $(\pi_{1:h}, e)$, we start from the parent frontier of the prefix $\pi_{1:h}$ and apply exactly one join corresponding to $e$,  so the prefix frontier serves as reusable state for all of its next-hop candidates.
The result of this one-step extension is a new frontier table for the longer prefix $\pi_{1:h+1} = \pi_{1:h} \circ e$. As before, we keep one row in the frontier table per distinct row reachable in the terminal table for each \texttt{(SeedId, timestamp)} pair. The new frontier table always updates \texttt{last\_id} to the identifier of the newly reached terminal row. Per our design, the new frontier table also carries the foreign-key columns required for future joins. Listing~\ref{lst:frontier-extend-rev-abstract} shows the query for
expanding a frontier with a reverse join. Each matching row in the
next table contributes a unique \texttt{last\_id}, so the query does
not require \texttt{DISTINCT}. A forward join, shown in Listing~\ref{lst:frontier-extend-fwd-abstract}, uses a similar query with a different join condition, but requires \texttt{SELECT DISTINCT} because multiple foreign-key values may map to the same primary key in the next table.

\begin{lstlisting}[style=sqlpretty, caption={Extending a frontier by a reverse join.}, label={lst:frontier-extend-rev-abstract}]
CREATE TEMP TABLE frontier_(*@\meta{extended\_prefix}@*) AS
SELECT
    SeedId,
    timestamp,
    next_table.primary_key AS last_id,
    (*@\meta{required\_foreign\_key\_1}@*),
    (*@$\ldots$@*),
    (*@\meta{required\_foreign\_key\_m}@*)
FROM frontier_(*@\meta{prefix}@*)
JOIN next_table
  ON next_table.foreign_key = frontier_(*@\meta{prefix}@*).last_id
WHERE (*@\meta{temporal\_constraint}@*);
\end{lstlisting}

\begin{lstlisting}[style=sqlpretty, caption={Extending a frontier by a forward join.}, label={lst:frontier-extend-fwd-abstract}]
CREATE TEMP TABLE frontier_(*@\meta{extended\_prefix}@*) AS
SELECT DISTINCT
    SeedId,
    timestamp,
    next_table.primary_key AS last_id,
    (*@\meta{required\_foreign\_key\_1}@*),
    (*@$\ldots$@*),
    (*@\meta{required\_foreign\_key\_m}@*)
FROM frontier_(*@\meta{prefix}@*)
JOIN next_table
  ON next_table.primary_key = frontier_(*@\meta{prefix}@*).foreign_key
WHERE (*@\meta{temporal\_constraint}@*);
\end{lstlisting}

\vspace*{-4pt}
\subsection{Metapath Statistics: log-count and log-rate}
\label{sec:metapath_statistics}

A frontier table tells us which rows are reachable by a given prefix
for a given seed node. As discussed in Section \ref{sec:metapath_informativenss}, variation in this occurrence pattern across seed nodes can provide predictive signal. To evaluate candidate extensions, \sysname{} summarizes the occurrence pattern using two complementary statistics: \emph{log-count}, which captures the total amount of distinct evidence reached by a metapath, and \emph{log-rate}, which captures the expansion introduced by a candidate extension relative to its parent prefix frontier.

These summaries are related to prior work on heterogeneous graphs and relational retrieval, where metapaths are treated as relation sequences with distinct semantics and path-based summaries are used to capture similarity or relevance~\cite{pathsim,hetesim,lao2010relational}. \sysname{} follows the same intuition but uses lightweight SQL statistics over frontier tables instead of complex similarity scores or random walks.

The first statistic is \emph{log-count}, which measures the total amount of distinct evidence reached by a metapath. Recall that, for a labeled training seed node $(s,t,y)$ and a metapath $\pi_{1:h}$, $N_h(s,t,\pi_{1:h})$ denotes the number of distinct terminal rows in the hop-$h$ frontier table. We define
\[
\mathrm{log\text{-}count}_h(s,t,\pi_{1:h})
=
\ln\!\bigl(1 + N_h(s,t,\pi_{1:h})\bigr).
\]
Log-count therefore captures the absolute volume of distinct reachable evidence. We apply $\ln(1+\cdot)$ to flatten the potentially heavy-tailed distribution of $N_h(s,t,\pi_{1:h})$.

The second statistic is \emph{log-rate}, which measures how strongly a candidate extension expands relative to the size of its parent frontier. For a candidate extension $(\pi_{1:h},e)$, let the extended metapath be $\pi_{1:h+1}=\pi_{1:h}\circ e$. We define
\[
\mathrm{log\text{-}rate}_{h+1}(s,t,\pi_{1:h+1})
=
\ln\!\left(
1 +
\frac{N_{h+1}(s,t,\pi_{1:h+1})}
     {N_h(s,t,\pi_{1:h})}
\right).
\]
Here, $N_{h+1}(s,t,\pi_{1:h+1})$ is the number of distinct rows reached after applying the candidate extension, and $N_h(s,t,\pi_{1:h})$ is the number of distinct rows already reached by its parent prefix. Log-rate therefore captures the extension's relative expansion by measuring how much evidence it reaches per row already present in the parent frontier.

Log-count and log-rate capture complementary aspects of an extension's occurrence pattern. For some extensions, the absolute number of distinct rows reached varies systematically with the task label. For others, the more informative pattern is how much the extension expands relative to its parent frontier, even when the absolute number of reached rows is modest. Log-count captures the first type of variation, while log-rate captures the second. Considering both allows \sysname{} to identify predictive occurrence patterns through either absolute reach or relative expansion.

\begin{figure}[t]
    \centering
    \includegraphics[width=\columnwidth]{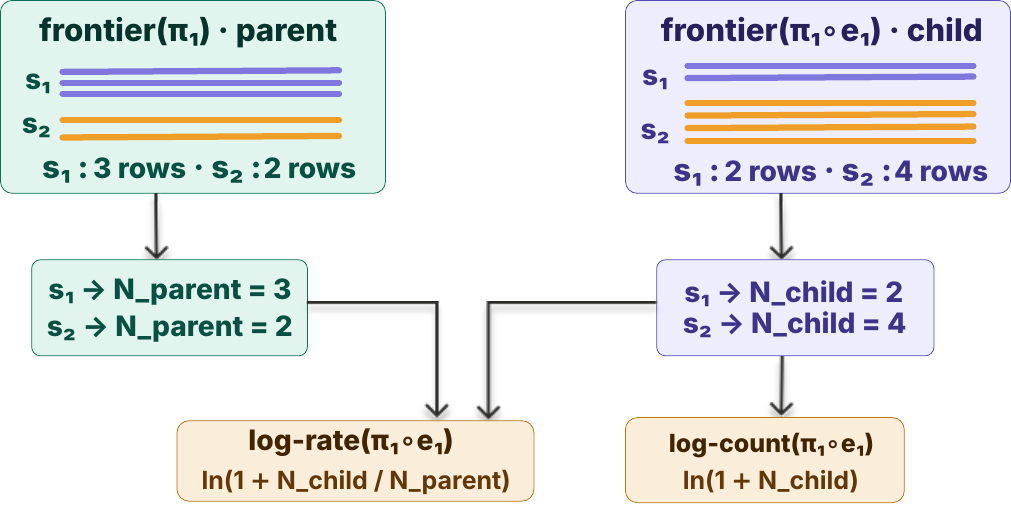}
    \caption{Reduction of frontier tables to metapath statistics. In this example, $s_1$ and $s_2$ denote two seeds, each identified by \texttt{(SeedId, timestamp)}. Each colored bar represents one distinct reachable row belonging to $s_1$ or $s_2$. By grouping the frontier rows by seed and applying \texttt{COUNT(*)}, we obtain the parent and child count used for log-count and log-rate.}
    \label{fig:frontier_reduction}
    \vspace*{-14pt}
\end{figure}

Figure~\ref{fig:frontier_reduction} illustrates how a parent frontier and its child frontier are reduced to per-seed metapath statistics. For each seed identified by \texttt{(SeedId, timestamp)}, we group the frontier rows using \texttt{GROUP BY} and apply \texttt{COUNT(*)} to obtain the frontier size. Applying this operation to the child frontier gives the count used for log-count, while applying it to both the parent and child frontiers gives the counts used for log-rate.

Because log-rate for a candidate extension depends on both the parent and child frontiers, it can only be computed starting at hop 2. Consequently, pruning begins with hop-1 prefixes and their candidate extensions at hop 2, and no pruning is applied directly at hop 1. Frontier tables are retained only while they are needed for further expansion and aggregation. For example, computing log-rate at hop $h+1$ requires access to both the parent frontier at hop $h$ and the child frontier at hop $h+1$. After the required statistics have been computed, the corresponding frontier tables can be dropped, limiting unnecessary intermediate storage.

\section{Batchwise Evaluation of Statistics}
\label{sec:batchwise_protocol}
The previous section treats all seed nodes in the training set $\mathcal{S}_{\mathrm{train}}$ as one batch for statistics computation.
\sysname{} actually evaluates candidate extensions on multiple disjoint subsets of training seed nodes (i.e., multiple batches of seed nodes). The log-count and log-rate statistics are computed separately for each batch, and the resulting distribution of statistics lets us identify stable, repeatable signals for metapath pruning. Batching also enables us to address label imbalance while forming the batches.

\subsection{Seed Node Batching and Label-aware Subsampling}
We score candidate extensions on a set of labeled training seed nodes \(\mathcal{S}_{\mathrm{eval}} \subseteq \mathcal{S}_{\mathrm{train}}\). In many cases, \(\mathcal{S}_{\mathrm{eval}} = \mathcal{S}_{\mathrm{train}}\), but when the training set is very large or highly imbalanced, we form \(\mathcal{S}_{\mathrm{eval}}\) by subsampling a smaller label-aware subset. This reduces the cost of SQL-based scoring while preventing the evaluation from being dominated by a majority label, which could obscure rarer but important signals~\cite{he2009learning, king2001logistic}. We implement this by capping the number of seed nodes per label. Although related in spirit to class-aware resampling~\cite{he2009learning,chawla2002smote}, the goal here is not to rebalance for model training, but to construct an informative subset for candidate scoring.

After this optional subsampling step, we partition $\mathcal{S}_{\mathrm{eval}}$ into $B$ approximately equal-sized batches, $\mathcal{S}_{\mathrm{eval}} = \bigcup_{b=1}^{B} \mathcal{S}_b$. Each batch $\mathcal{S}_b$ contains labeled training seed nodes and serves as the input for frontier materialization and candidate scoring (Section \ref{sec:scoring}). We implement deterministic batching using the SQL window function \texttt{NTILE}, which first orders the seed nodes by identifier and then assigns them to $B$ buckets of approximately equal size (Listing~\ref{lst:seed_division}). 

\begin{lstlisting}[style=sqlpretty, caption={Assigning training seed nodes to batches.}, label={lst:seed_division}]
CREATE TABLE seed_batches AS
SELECT
  SeedId,
  timestamp,
  label,
  NTILE(TARGET_N_BATCHES) OVER (ORDER BY SeedId) AS batch_id
FROM seeds_eval;
\end{lstlisting}

\subsection{Benefits of Batchwise Scoring}
\label{sec:why_batchwise}
We compute the metapath statistics log-count and log-rate for all $(\pi_{1:h},e)$ pairs separately on each batch, and we use the batchwise statistics to score the metapaths. This has two benefits:
\begin{itemize}
\item \textbf{Stability.} Candidate extensions that score well across many batches are more likely to reflect a repeatable relationship between the statistical evidence and the label. Candidates that score well in only a few batches are treated as less reliable, and the spread of scores across batches reveals whether a candidate extension is informative for a broad subset of training seed nodes~\cite{stability}.

\item \textbf{Scalability.} Because the batches are disjoint, their statistics can be evaluated concurrently and independently by multiple worker threads.
Each worker thread can process one batch end-to-end, including frontier materialization, metapath statistics generation, and scoring, without any cross-worker synchronization. This parallel execution can be implemented on any DBMS 
and results in throughput that scales linearly with the number of workers due to the coordination-free nature of the parallelism.
\end{itemize}

%% file: sections/scoring.tex
\section{Scoring Candidate Extensions}
\label{sec:scoring}
After the SQL-based computation of the log-count and log-rate statistics for the different batches and metapaths, what remains is scoring the candidate extensions based on these statistics, which happens outside the DBMS.
Our scoring depends on the utility of an extension to learning the training labels, its cost, and the frequency of the extension in the database. We discuss the details next.

\subsection{Mutual Information as a Proxy for Utility}
\label{sec:mi_proxy}

Each training seed node has an associated task-specific label, and our goal is to retain candidate extensions whose statistics are predictive of that label while pruning extensions that are statistically uninformative.
To evaluate this, we treat the log-count and log-rate values for an extension as candidate features $X$, the task label as the target $Y$, and measure their dependence using mutual information (MI) $I(X;Y)$. This provides a concrete measure of the predictive signal discussed in Section \ref{sec:metapath_informativenss}, where an extension is informative when its log-count or log-rate values vary systematically with the task label. If these statistics are largely independent of the label, then the extension provides little occurrence-based predictive signal. This use of mutual information is consistent with prior work showing that path-based relational summaries can be informative for prediction~\cite{pathsim,hetesim,lao2010relational}.

MI is commonly used in supervised feature selection because it quantifies feature-target dependence without requiring a specific predictive model~\cite{battiti1994mi,peng2005feature}. In our setting, this makes MI a natural task-specific way to assess whether a log-count or log-rate statistic carries useful signal for the corresponding prediction task. MI is also model-agnostic, captures potentially non-linear associations, and is zero if and only if $X$ and $Y$ are independent.

We use MI as a cheap proxy for the utility of an extension because directly estimating utility by training a predictive model and monitoring validation error would be prohibitively expensive. For a candidate extension $(\pi_{1:h},e)$, we would need to train a model using the per-seed-node statistics available up to $\pi_{1:h}$ and obtain its validation error $E_{\pi_{1:h}}$. We would also train a model on the extended metapath $\pi_{1:h+1}=\pi_{1:h}\circ e$ using the union of the prefix statistics and the new ones introduced by the extension. If the validation error of that model is $E_{\pi_{1:h+1}}$, the utility of candidate extension $(\pi_{1:h},e)$ can be computed as $E_{\pi_{1:h}}-E_{\pi_{1:h+1}}$.
This procedure would require training many models and is not practical for large relational databases with complex schemas.

\subsection{Estimating Mutual Information}
\label{sec:mi_estimation}

In our setting, MI is not available in closed form since we only observe
seed nodes, task labels, and metapath statistics for each batch. We therefore estimate MI from data using standard nonparametric estimators based on $k$-nearest-neighbor distances.

We estimate MI separately for log-count and log-rate within each batch of seed nodes.
For batch $b$, candidate extension $(\pi_{1:h}, e)$, and statistic $g\in\{\text{log-count},\text{log-rate}\}$, let $X_b(\pi_{1:h}, e, g)$ denote the vector of values of the statistic $g$ for the seed nodes in $b$. Let $Y_b$ denote the corresponding vector of task labels. We estimate a batchwise mutual information between $X_b$ and $Y_b$, denoted by $MI_b(\pi_{1:h}, e, g)$. For regression tasks, both $X_b$ and $Y_b$ are continuous, so we use the Kraskov-St{\"o}gbauer-Grassberger (KSG) estimator~\cite{kraskov2004mi}, which estimates MI from nearest-neighbor distances and gives higher values when similar feature values tend to correspond to similar label values. For classification tasks, the statistic (log-count or log-rate) is continuous while the task label is discrete, so we use Ross's mixed discrete-continuous estimator~\cite{ross2014mi}. This estimator applies the same nearest-neighbor idea to the mixed continuous-discrete setting and gives higher values when feature values are more similar within the same class than across different classes.
These standard estimators are readily available in libraries such as scikit-learn~\cite{scikitlearn}.

\subsection{Entropy-normalized Mutual Information}
\label{sec:nmi}

The batchwise MI estimates introduced above are not directly comparable across batches for a given extension because the uncertainty of the task labels can vary from one batch to another. For example, in a classification task, one batch may have a more balanced label distribution than another, while in a regression task, one batch may span a wider range of target values than another. As a result, the same raw MI value can represent different levels of relevance depending on how much label uncertainty is present in that batch. To make MI values comparable across batches, we normalize each batchwise MI by an entropy term, so that the resulting score measures the fraction of label uncertainty explained by the statistic rather than its absolute MI magnitude. This follows the standard information-theoretic view that normalization is needed when MI values are compared across settings with different entropies~\cite{vinh2010ami}.

We use the empirical entropy of the labels in each batch $b$, computed as
$
H(Y_b)
=
-\sum_{y} p_b(y)\,\ln p_b(y)
$.
The summation is taken differently for classification and regression tasks.
For classification tasks, we sum over the different label values $y$ in batch $b$. Thus, 
$p_b(y)$ is the frequency of label value $y$ in batch $b$ normalized by the size of $b$.
For regression tasks, the task label is continuous, so we first discretize it into bins. We choose the number of bins and their ranges once from the training data and then use the same bin count for all batches. 
We use quantile-based binning so that the bins remain populated even for skewed or heavy-tailed targets~\cite{fayyad1993multi}.
After discretization, entropy is computed by taking the summation over the bins. That is, $y$ ranges over the bins and $p_b(y)$ is the number of times bin $y$ appears in the batch divided by the batch size.

We define the entropy-normalized MI score for batch $b$ as
\vspace*{-4pt}
\[
\mathrm{NMI}_b (\pi_{1:h}, e, g)
=
\frac{MI_b(\pi_{1:h}, e, g)}{H(Y_b)}.
\]
\vspace*{-2pt}

It is possible for $H(Y_b)$ to be zero. This happens when the batch contains no variation in the task label, for example, a batch in a classification task where all seed nodes belong to the same class. 
In this case, we set $\mathrm{NMI}_b$ for this batch to zero.

\subsection{Cost-aware Scoring}
\label{sec:cost_aware_scoring}

For a given batch, entropy-normalized MI measures how strongly a candidate extension is associated with the task label, but it does not account for how expensive that extension is to sample. In GNN training, computation and memory grow with the size of the sampled neighborhood, and high-fanout extensions can quickly lead to neighborhood explosion~\cite{graphsage,graphsaint,labor}. We therefore score each candidate extension in a batch using not only its $\mathrm{NMI}_b$, but also the \emph{cost} of sampling it, as measured by the number of paths it introduces.

We derive the cost of a candidate extension from its log-count values. For a given candidate extension $(\pi_{1:h},e)$ and a given batch $b$, the cost is defined as
\[
\mathrm{cost}_b(\pi_{1:h},e)
=
\mathrm{avg}_{s_i \in b}\; \bigl(e^{\mathrm{log\text{-}count}_h(s_i,t_i,\pi_{1:h})} - 1\bigr)
\]
where $s_i$ is a seed node in batch $b$ and $t_i$ is the corresponding timestamp. The expression within parentheses recovers, for the seed node $s_i$, the actual count of paths for the extension by inverting the equation of $\mathrm{log\text{-}count}$. The costs of the extension is then computed as the average of all counts in the batch.
Thus, an extension has low cost when it reaches a small number of rows on average and high cost when it creates large expansions.

We combine relevance, as measured by $\mathrm{NMI}_b$, and cost into a single batchwise score.
For candidate extension $(\pi_{1:h},e)$, statistic $g \in \{\text{log-count}, \text{log-rate}\}$, and batch $b$, we define
\[
\mathrm{score}_b(\pi_{1:h},e,g)
=
\frac{\mathrm{NMI}_b(\pi_{1:h},e,g)}{\mathrm{cost}_b(\pi_{1:h},e)}.
\]
Thus, an extension receives a high score only when it is both label-informative and cheap to sample within the batch, and 
\sysname{} prunes metapaths with low score values.
This follows the same principle as 
cost-sensitive feature selection, where a feature is preferred when it is informative yet inexpensive to acquire~\cite{turney1994cost,greiner2002learning,controllingcost,featurebudget}.

\subsection{Coverage as a Measure of Support}
\label{sec:coverage}

The scores computed in the previous section tell us whether an extension is informative for the task label and whether it is cheap to sample, but they do not tell us how broadly that extension is observed across the training seed nodes.
That is, an extension may receive a high score because it behaves well on a small subset of training seed nodes, yet still be observed too infrequently to support a reliable pruning decision. We therefore introduce a third statistic for evaluating extensions, which we call \emph{coverage}. Unlike log-count and log-rate, coverage is label-agnostic. Its role is to measure how frequently an extension is present in the training data.
This is analogous to the notion of \emph{support} in frequent itemset data mining.

For a candidate extension $(\pi_{1:h},e)$ and a batch $b$,
coverage is the fraction of training seed nodes for which the extended frontier is non-empty,
and it is computed by recovering $N_{h+1}$ from log-count:
\[
\mathrm{coverage}_{b}(\pi_{1:h},e)
=
\frac{1}{|b|}
\sum_{s_i \in b}
\mathbb{I} \bigl (
N_{h+1}(s_i,t_i,\pi_{1:h}\circ e) > 0
\bigr ).
\]
where $s_i$ is a seed node in $b$, $N_{h+1}$ is the count of the candidate extension, $\mathbb{I}(\cdot)$ is the indicator function which returns 1 if its argument is true and 0 otherwise. A zero $N_{h+1}$ value indicates that no row was reached for this extension and this seed node, and too many zero $N_{h+1}$ values indicate a low-support metapath.

\subsection{Combining the Statistics into a Quality Score}
\label{sec:q_score}

\begin{sloppypar}
For every candidate extension $(\pi_{1:h},e)$, we compute for every batch $b=1 \ldots B$ the score for log-count, the score for log-rate, and the coverage. Across the $B$ batches, these values form three \emph{empirical distributions}, which must be combined into a single quality score to determine which metapaths to sample and which to prune.
\end{sloppypar}

Our approach combines these scores in two steps. First, we summarize each of the three distributions as a single number. Next, we combine the three summary numbers into one overall quality score per candidate extension.

\vspace*{4pt}
\noindent\textbf{Summarizing Each Distribution.}
A natural way to summarize the distribution of a statistic is to take the mean. We instead adopt a more conservative summary: a one-sided \emph{lower confidence bound} (LCB) on the mean. The LCB discounts extensions whose performance appears good only because of a few favorable batches. In this respect, it plays a role analogous to stability-oriented selection criteria, which favor signals that remain strong under repeated subsampling over signals that are only occasionally strong~\cite{stability}.

For any batchwise statistic $z_1, \ldots, z_B$, under a Gaussian i.i.d.\ assumption, the LCB is defined as
\[
\mathrm{LCB}_\delta\!\left(\{z_i\}_{i=1}^{B}\right)
= \hat{\mu} - t_{1-\delta,\,B-1} \, \frac{\hat{\sigma}}{\sqrt{B}},
\]
where $\hat{\mu}$ and $\hat{\sigma}$ are the sample mean and standard deviation across batches, and $t_{1-\delta,\,B-1}$ is the $(1-\delta)$ quantile of the Student-$t$ distribution with $B-1$ degrees of freedom. We use the same significance level $\delta$ for all three statistics (log-count, log-rate, and coverage). The parameter $\delta$ controls how conservative the resulting estimate is, and we tune it for each specific prediction task.

For every candidate extension $(\pi_{1:h},e)$, 
we compute an LCB for log-count, log-rate, and coverage, e.g.,
the formula for log-count is 
$\mathrm{LCB}_\delta\!\left(\left\{\mathrm{score}_b(\pi_{1:h},e,\text{log-count})\right\}_{b=1}^{B}\right)$. We remove extensions for which the LCB of coverage is zero, since this occurs only when the extension does not lead to any actual paths in the data.

The remaining candidate extensions are grouped by hop, and percentile ranks are computed within each hop. Hop-wise grouping matters because the numerical range of the LCB values shifts with hop depth: paths at deeper hops typically exhibit different fanout behavior, span different attainable count scales, and induce different relevance distributions than shallower ones, making raw LCB values incomparable across hops. To eliminate this scale mismatch while preserving the relative ordering of candidates within each hop, we sort the metapaths at each hop separately by each of the three LCB values and compute the percentile rank of each candidate within each sorted list. Thus, each candidate extension $(\pi_{1:h},e)$ now has three scores on the common scale of $[0,1]$ summarizing the distribution of each of its three statistics as a single number:
\[
p_{\mathrm{log-count}}(\pi_{1:h},e)
=
\mathrm{rank}_\%\!\left(\mathrm{LCB}_\delta\!\left(\left\{\mathrm{score}_b(\pi_{1:h},e,\text{log-count})\right\}_{b=1}^{B}\right)\right),
\]
\[
p_{\mathrm{log-rate}}(\pi_{1:h},e)
=
\mathrm{rank}_\%\!\left(\mathrm{LCB}_\delta\!\left(\left\{\mathrm{score}_b(\pi_{1:h},e,\text{log-rate})\right\}_{b=1}^{B}\right)\right),
\]
\[
p_{\mathrm{coverage}}(\pi_{1:h},e)
=
\mathrm{rank}_\%\!\left(\mathrm{LCB}_\delta\!\left(\left\{\mathrm{coverage}_b(\pi_{1:h},e)\right\}_{b=1}^{B}\right)\right).
\]

\vspace{4pt}
\noindent\textbf{Computing One Overall Quality Score.}
As discussed in Section 6.2, a candidate extension can be useful for different reasons. Some extensions are best characterized by the total amount of distinct evidence they expose, which is captured by log-count. Others are better characterized by how strongly they expand relative to the parent frontier, which is captured by log-rate. In practice, an extension is often well represented by one of these two label-dependent statistics, but not necessarily both. We therefore represent the relevance of a candidate extension by the stronger of log-count and log-rate. That is, we compute
\[
p_{\max}(\pi_{1:h},e)
=
\max\!\left(p_{\text{log-count}}(\pi_{1:h},e),\, p_{\text{log-rate}}(\pi_{1:h},e)\right).
\]
This avoids penalizing an extension simply because one way of looking at relevance is less appropriate for its semantics.

We also include coverage in the final score of the candidate extension. Coverage is a measure of support and not a measure of feature-label dependence, but it remains important because even a label-informative extension should not be preferred if it is only present in a small fraction of training seed nodes. Our final quality score for a candidate extension is therefore
\[
Q(\pi_{1:h},e)
=
p_{\max}(\pi_{1:h},e)
+
p_{\mathrm{coverage}}(\pi_{1:h},e).
\]

\subsection{Separating Promising and Poor Extensions}
\label{sec:thresholding}

\sysname{} should keep candidate extensions with a high quality score  $Q(\pi_{1:h},e)$ and prune ones with a low quality score. The question is where to draw the line separating good and bad extensions.

We develop a data-driven process to make this decision. We standardize the $Q$ values by subtracting the mean and dividing by the standard deviation. We then use expectation-maximization \cite{mclachlan2000finite} to fit a two-component Gaussian mixture model to the resulting $Q$ values. The component with the larger fitted mean is labeled \emph{good}, and the component with the smaller fitted mean is labeled \emph{bad}.

Because $Q$ is based on $p_{\max}$ and $p_{\mathrm{coverage}}$, which are hop-wise percentile ranks, the $Q$ values are much better aligned across hops than the raw LCB values. This makes it possible to apply a single global mixture model to the standardized $Q$ scores while still respecting the different raw scales that arise at different hop depths.

\vspace*{-4pt}
\section{Generating Sampling Rules}
\label{sec:rule_generation}

After clustering the candidate extensions into good and bad clusters, the remaining task is to convert these cluster assignments into actionable pruning decisions for neighborhood sampling.

We generate a \emph{pruning rule} for each candidate extension that indicates whether the extension should be kept and expanded or pruned:
$
\mathcal{R}(\pi_{1:h},e)\in\{\texttt{expand},\texttt{prune}\}.
$
For a prefix $\pi_{1:h}$ and candidate extension $e$, let $\pi_{1:h}\circ e$ denote the corresponding extended metapath. We mark the extension as \texttt{expand} if the extended metapath is itself assigned to the good cluster, or if it has at least one descendant metapath assigned to the good cluster. Intuitively, even when the immediate extension is not highly ranked, it should be retained if it serves as a gateway to a strong downstream extension. Conversely, we mark the extension as \texttt{prune} only when the extended metapath and \emph{all} of its descendants lie in the bad cluster.

The pruning rules take the form: \emph{``after following prefix $\pi_{1:h}$, do not expand candidate $e$ at hop $h+1$.''} Because a hop $h$ node in the terminal table $T_h$ may be reachable through multiple schema-valid metapaths from a seed node, implementing these rules require the sampler to track not only the candidate being expanded at hop $h$ but also the prefix metapath that reached it.
Consider, for example, a candidate extension $e$ reachable through two prefix metapaths $\pi_{1:h}^{A}$ and $\pi_{1:h}^{B}$, with $\mathcal{R}(\pi_{1:h}^{A}, e) = \texttt{expand}$ and $\mathcal{R}(\pi_{1:h}^{B}, e) = \texttt{prune}$.
The decision whether to expand (i.e., sample neighbors) depends on which of the two prefix metapaths was traversed to reach $e$.

If the neighborhood sampler used in the GNN training framework tracks at runtime the prefix used to reach a node while sampling, the pruning rules can be implemented directly in the sampler. However, standard frameworks do not track prefixes while sampling. For example, \sysname{} uses the popular PyG library~\cite{pyg,pyg2}, whose sampler allows the user to specify the extensions to sample and the number of samples per extension type at each hop, but does not track the prefix used to reach each hop. As a result, it cannot distinguish between the two cases described above.

To bridge this gap, we approximate the prefix-aware pruning rules with prefix-agnostic rules of the form:
\emph{``if a node at hop $h$ belongs to the terminal table $T_h$ of the prefix metapath $\pi_{1:h}$ of a rule, do not expand candidate $e$ at hop $h+1$.''}
That is, ignore the specific prefix used to reach $T_h$.
To specify that a candidate not be expanded, we instruct PyG to sample zero neighbors for that candidate at hop $h$, leaving the default sampling unchanged for other extensions.

Under this approximation, if the terminal table of prefix $\pi_{1:h}$ is reachable through multiple prefixes and even one of them carries a rule $\mathcal{R}(\pi_{1:h}, e) = \texttt{prune}$, the extension $e$ is pruned. We also experimented with alternative approximations that aggregate the rules associated with all prefixes reaching the terminal table of $\pi_{1:h}$ --- for example, pruning only when a majority of these rules indicate pruning --- and found that our simple approximation works best.

%% file: sections/experiments.tex
\section{Experiments}
\label{sec:experiments}

\subsection{Experimental Setup}
\label{sec:expsetup}

\subsubsection{Datasets and Tasks}
\begin{sloppypar}
We evaluate \sysname{} on the RelBench benchmark, which provides realistic databases, temporally scoped predictive tasks, and standardized training/validation/test splits with task-specific metrics~\cite{relbench}.
Our evaluation focuses on \textbf{node-level classification and regression tasks}, which are the two settings supported by our metapath pruning pipeline. We use five RelBench databases with complex schemas and large instances:
\texttt{rel-ratebeer},
\texttt{rel-trial},
\texttt{rel-f1}, 
\texttt{rel-stack}, and
\texttt{rel-avito}. 
Table~\ref{tab:relbench_dataset_characteristics} summarizes their domains, schemas (number of tables and number of foreign-key links), and sizes (number of rows).

We select these databases because their schemas admit many valid metapaths and contain highly connected tables, making metapath selection both useful and computationally challenging. Simpler schemas, such as \texttt{rel-amazon}, provide few to no meaningful pruning choices. Together, the five databases span different application domains, schema structures, task semantics, database sizes, and connectivity patterns, providing a diverse and comprehensive evaluation of \sysname{}. The number of rows in the databases ranges from \textasciitilde{}100K for \texttt{rel-f1} to \textasciitilde{}25M for \texttt{rel-avito}.
The \texttt{rel-ratebeer} database was introduced in RelBench v2~\cite{relbenchv2} and is the largest of our test databases by data size, despite being second-largest by row count (\textasciitilde{}17M rows).
\end{sloppypar}


\begin{table}[t]
\centering
\small
\setlength{\tabcolsep}{3pt}
\renewcommand{\arraystretch}{1.15}
\resizebox{\columnwidth}{!}{%
\begin{tabular}{@{}llccc@{}}
\toprule
Database & Domain & Tables & FK Links & Rows \\
\midrule
rel-ratebeer & Beverage reviews   & 13 & 21 & 16,874,623 \\
rel-trial    & Clinical trials    & 15 & 15 & 5,852,157  \\
rel-f1       & Formula 1 racing   & 9  & 13 & 97,606     \\
rel-stack    & Online Q\&A        & 7  & 11 & 5,399,818  \\
rel-avito    & Online advertising & 8  & 11 & 24,653,915 \\
\bottomrule
\end{tabular}%
}
\caption{Characteristics of the evaluated RelBench databases. Tables is the number of tables in the database. FK Links denotes the number of foreign-key columns. Rows is the database size measured as the number of rows in all tables.}
\label{tab:relbench_dataset_characteristics}
\vspace*{-24pt}
\end{table}

\vspace*{-4pt}
\subsubsection{Hardware and Software Environment}

Our implementation uses DuckDB \cite{duckdb} v1.2.0 as the DBMS.
All \sysname{} workers run as threads within the same Python process, and each worker uses its own private, embedded DuckDB instance, with each instance having multiple query execution threads (8 threads per worker by default). 
The RelBench database is attached to each  DuckDB instance in read-only mode.
Each worker gets a memory limit of \(M_{\mathrm{worker}} = 0.65M_{\mathrm{available}}/W\), where \(M_{\mathrm{available}}\) is the available system memory and \(W\) is the number of workers.

We use Pytorch with the Pytorch Geometric (PyG) library v2.6.1 as the sampling and GNN framework. All experiments were conducted on a GPU cluster with H100 NVIDIA GPUs with 80\,GB of memory each and AMD CPUs. Each experiment uses 1 GPU and 2 TB of RAM. The NVIDIA driver is v570.172.08 and CUDA is v12.8.

\vspace*{-4pt}
\subsubsection{Sampling Strategies}
We compare \sysname{} to two other strategies for constructing the training subgraph for each seed node. 

\vspace{4pt}
\noindent
\textbf{Random $m$-hop sampling.} Starting from the seed node, expand the frontier for $m$ successive hops. At each hop, every frontier node draws a uniform sample of its neighbors with a fixed fanout applied identically across all edge types and hops~\cite{graphsage}. This is the default sampling strategy for GNN training, and it is metapath-agnostic: it makes no distinction among the relational paths it traverses.

\vspace{4pt}
\noindent
\textbf{MPS-GNN.} MPS-GNN~\cite{mpsgnn} is the state-of-the-art in metapath selection for RDL. We use the implementation provided by the authors, adapting it to the much larger RelBench databases. MPS-GNN is designed for binary node classification and was originally evaluated on simpler databases without a temporal dimension. To scale MPS-GNN to the more complex and temporal RelBench databases, we first draw a small label-balanced subset of training seed nodes, together with a small validation subset, and then construct a single temporally consistent $H$-hop sampled subgraph around those seeds. We pass this sampled subgraph, together with the classification labels of the seeds, to the MPS-GNN search routine. The search is constrained by two parameters: $k$, which keeps only the top-$k$ candidate extensions at each step, and \texttt{beam\_width}, which caps the number of partial metapaths retained after each expansion round. These constraints are essential in practice, since increasing either parameter quickly expands the number of explored metapaths, leading to a sharp increase in search time even for small databases. We use $k=3$ and \texttt{beam\_width}=6 by default. The metapaths evaluated by this search are ranked, and our default is to retain the top 20 and prune the rest, ensuring sufficient data coverage. If MPS-GNN evaluates fewer than 20 metapaths, we retain them all. We also investigate the effect of evaluating and retaining more metapaths in Section~\ref{sec:sensitivity_mpsgnn}.

\vspace*{-4pt}
\subsubsection{GNN Backbones}
The sampling strategies can be applied to any GNN backbone. We evaluate them with three backbone models that represent message-passing and transformer-based architectures.
These are also the backbones used in~\cite{relgt}.

\vspace{4pt}
\noindent
\textbf{HeteroGraphSAGE.}
This is a heterogeneous version of the GraphSAGE~\cite{graphsage} message-passing GNN, and it is the default backbone in the RelBench~\cite{relbench} implementation. It extends GraphSAGE to multiple node and edge types by performing neighbor aggregation per edge type and combining the resulting messages to form the updated embedding. Thus, each edge type contributes through its own transformation before being merged into a single representation.

\vspace{4pt}
\noindent
\textbf{HGT.}
The Heterogeneous Graph Transformer (HGT)~\cite{hgt} replaces fixed neighborhood aggregation with attention-based message passing. HGT learns separate transformations for different node types when computing attention, and separate relation-specific parameters for different edge types when passing messages. This means that, when updating a node, the model can assign different importance to neighbors depending on both their node type and the type of relation connecting them. As a result, HGT can model heterogeneous structure more flexibly than HeteroGraphSAGE while still operating as a message-passing model.

\vspace{4pt}
\noindent
\textbf{RelGT.}
The Relational Graph Transformer (RelGT)~\cite{relgt} is a recently introduced state-of-the-art backbone specifically designed for RDL, and it shows superior accuracy for RelBench tasks. For each seed node, RelGT samples a local subgraph and selects a fixed set of $k$ neighboring nodes to form the input token set. Each selected node is represented by a multi-element token encoding its features, type, hop distance, time, and local structure. RelGT then combines local self-attention over these sampled tokens with global attention via learnable centroid tokens, allowing it to capture both neighborhood-level and database-wide patterns.

\vspace*{-8pt}
\subsubsection{Hyperparameters.}
We train the GNNs for 10, 20, or 30 epochs, depending on the dataset size and the GNN complexity. For a given dataset, each GNN backbone is trained for the same number of epochs across all sampling strategies, ensuring a fair comparison among the strategies. For experiments configured to run more than 10 epochs, we apply early stopping based on improvement in the validation metric. The neighborhood sampling fanout is set to 64 at every hop in all experiments, except for RelGT on large tasks such as \texttt{beer-churn} and \texttt{post-votes}, where we use a fanout of 16 to control training cost. The number of GNN layers is always matched to the number of sampled hops for message passing models.

In \sysname{}, the $\delta$ hyperparameter controls the strength of the uncertainty penalty in the LCB estimates used to aggregate log-count, log-rate, and coverage. Smaller $\delta$ values produce a larger subtraction term in the LCB formula and hence a more conservative estimate that more strongly penalizes extensions with high batch-to-batch variability. The $\delta$ value has to be below $0.5$, so for each experiment we choose the best $\delta$ from $\{0.1, 0.2, 0.3, 0.4\}$ based on the best validation accuracy, using time per epoch as the secondary criterion when the validation accuracies are close.

\begin{table*}[t]
\centering
\small
\setlength{\tabcolsep}{3.5pt}
\renewcommand{\arraystretch}{1.15}
\begin{tabular}{ll l cc cc cc}
\toprule
\multirow{2}{*}{Dataset} &
\multirow{2}{*}{Task Name} &
\multirow{2}{*}{GNN Architecture} &
\multicolumn{2}{c}{Random} &
\multicolumn{2}{c}{MPS-GNN} &
\multicolumn{2}{c}{\sysname{} (ours)} \\
\cmidrule(lr){4-5}\cmidrule(lr){6-7}\cmidrule(lr){8-9}
& & &
Test $\uparrow$ & Time/epoch $\downarrow$ &
Test $\uparrow$ & Time/epoch $\downarrow$ &
Test $\uparrow$ & Time/epoch $\downarrow$ \\
\midrule
\multirow{3}{*}{rel-ratebeer} &
\multirow{3}{*}{user-churn} &
HeteroGraphSAGE (3 hop) & 0.923 & 1,397 & 0.949 & 332 & \textbf{0.951} & \besttime{160} \\
& & HGT (3 hop)          & 0.875 & 3,781 & \textbf{0.894} & 1,010 & 0.875 & \besttime{721} \\
& & RelGT (3 hop, k=50)         & 0.933 & 9,742 & 0.942 & 2,443 & \textbf{0.953} & \besttime{1,253} \\
\midrule
\multirow{3}{*}{rel-ratebeer} &
\multirow{3}{*}{beer-churn} &
HeteroGraphSAGE (3 hop) & 0.784 & 3,103 & \textbf{0.814} & 1,554 & 0.785 & \besttime{518}\\
& & HGT (3 hop)           & \textbf{0.740} & 11,181 & 0.719 & 5,823 & 0.725 & \besttime{1,299} \\
& & RelGT (3 hop, k=50)         & 0.777 & 17,644 & 0.780 & 11,102 & \textbf{0.799} & \besttime{1,494} \\
\midrule
\multirow{3}{*}{rel-trial} &
\multirow{3}{*}{study-outcome} &
HeteroGraphSAGE (3 hop) & 0.648 & 16 & \textbf{0.668} & 8 & 0.660 & \besttime{5} \\
& & HGT (3 hop)           & 0.552 & 22 & \textbf{0.597}  & 16 & 0.569 & \besttime{11} \\
& & RelGT (3 hop, k=100)        & 0.691 & 146 & 0.676 &  119 & \textbf{0.702} & \besttime{97} \\
\midrule
\multirow{3}{*}{rel-f1} &
\multirow{3}{*}{driver-dnf} &
HeteroGraphSAGE (3 hop) & 0.712 & 27 & 0.676 & 17 & \textbf{0.723} & \besttime{6} \\
& & HGT (3 hop)           & 0.661 & 41 & 0.508 & 26 & \textbf{0.672} & \besttime{11} \\
& & RelGT (3 hop, k=100)        & 0.675 & 624 & 0.681 & 331 & \textbf{0.736} & \besttime{115}  \\
\midrule
\multirow{3}{*}{rel-f1} &
\multirow{3}{*}{driver-top3} &
 HeteroGraphSAGE (3 hop)          & \textbf{0.797} & 6 & 0.774 & 5 & 0.779 & \besttime{2} \\
& & HGT (3 hop)    & 0.666 & 9 & \textbf{0.695} & 8 & 0.673 & \besttime{4}\\
& & RelGT  (3 hop, k=100)       & 0.681 & 141 & 0.750 & 123 & \textbf{0.766} & \besttime{29} \\
\midrule
\multirow{3}{*}{rel-stack} &
\multirow{3}{*}{user-badge} &
HeteroGraphSAGE (4 hop) & \textbf{0.891} & 699 & 0.885 & 314 & 0.889 & \besttime{185} \\
& & HGT (4 hop)          & \textbf{0.854} & 818 & 0.854 & 425 & 0.852 & \besttime{234} \\
& & RelGT (3 hop, k=50)         & \textbf{0.878} & 3,808 & 0.877 & \besttime{2,720} & 0.877 & 3,063 \\
\midrule
\multirow{3}{*}{rel-stack} &
\multirow{3}{*}{user-engagement} &
HeteroGraphSAGE (4 hop) & \textbf{0.907} & 1,434 & 0.904 & \besttime{182} & 0.906 & 209 \\
& & HGT (4 hop)           & 0.851 & 1,846 & 0.845 & \besttime{221} & \textbf{0.876} & 248 \\
& & RelGT (3 hop, k=50)         & 0.904 & 4,636 &  0.901 & \besttime{3,368} & \textbf{0.905} & 3,697 \\
\midrule
\multirow{3}{*}{rel-avito} &
\multirow{3}{*}{user-visits} &
HeteroGraphSAGE (3 hop) & 0.660 & 52 & \textbf{0.662} & 35 & 0.661 & \besttime{18} \\
&  & HGT (3 hop)          & 0.626 & 70 & 0.632 & 45 & \textbf{0.635} & \besttime{25} \\
& & RelGT (3 hop, k=100)        & 0.654 & 924 & 0.653 & 663 & \textbf{0.661} & \besttime{427} \\
\midrule
\multirow{3}{*}{rel-avito} &
\multirow{3}{*}{user-clicks} &
HeteroGraphSAGE (3 hop) & \textbf{0.678} & 68 & 0.669 & 30 & 0.669 & \besttime{16} \\
& & HGT (3 hop)          & 0.648 & 55 & 0.633 & 33 & \textbf{0.660} & \besttime{17} \\
& & RelGT (3 hop, k=50)        & 0.649 & 742 & 0.640 & 565 & \textbf{0.670} & \besttime{325}  \\
\bottomrule
\end{tabular}
\caption{Node classification performance. Evaluation metric is AUCROC (higher is better). Epoch time is in seconds (lower is better). The fastest time/epoch is highlighted in green, and the best test metrics are in bold.}
\label{tab:node_classification_performance}
\vspace*{-16pt}
\end{table*}

\subsubsection{Summary of Experiments.}
Section~\ref{sec:improvement} shows that \sysname{} consistently improves training time for all GNN backbones. We also show that this improvement in training time does not come at the cost of reduced accuracy. On the contrary, the most accurate model overall is the powerful RelGT backbone with \sysname{} for metapath selection.
Section~\ref{sec:preprocessing} shows that \sysname{} preprocessing time is very reasonable, while MPS-GNN is prohibitively expensive.

The remaining experiments focus on sensitivity analysis and ablations.
In Section~\ref{sec:sensitivity_mpsgnn}, we show that retaining more of the metapaths selected by MPS-GNN (30 instead of 20) requires higher preprocessing time and results in higher epoch times during training, while not adding substantially to accuracy. Thus, our choice of retaining 20 metapaths for MPS-GNN is well-justified.

Next, we turn our attention to the internals of \sysname{}.
In Section~\ref{sec:ablation}, we show ablations over the scoring function and demonstrate that all of its components are necessary. We study the process of tuning the hyperparameter $\delta$ in Section~\ref{sec:delta-tuning}, and we show that this hyperparameter has an effect on accuracy and time per epoch, but \sysname{} is not overly sensitive to the choice of $\delta$ to the point of requiring unrealistic effort to tune this hyperparameter. Section~\ref{sec:prefix-agnostic-rule-approximation} compares the prefix-agnostic heuristic we use to generate sampling rules (Section~\ref{sec:rule_generation}) with prefix-aware versions and shows that the prefix-aware versions result in less metapath pruning and higher epoch times (as expected), but they do not add substantial accuracy, thus justifying our prefix-agnostic heuristic.

Finally, we end with a comprehensive scalability analysis of the SQL queries used by \sysname{} in Section~\ref{sec:scalability}. We show that the memory required to materialize the frontier grows linearly with frontier size.
However, this memory can be regulated by reducing the number of worker threads, increasing the number of seed-node batches (thus reducing batch size), and/or limiting the memory available to DuckDB and requiring it to spill to disk.
Regulating the memory consumed by SQL queries as the database grows is very well supported in a modern DBMS, and \sysname{} takes full advantage of this capability to ensure scalability.

\vspace*{-4pt}
\subsection{Improvement in GNN Training}
\label{sec:improvement}

\begin{table*}[t]
\centering
\small
\setlength{\tabcolsep}{3.5pt}
\renewcommand{\arraystretch}{1.15}
\begin{tabular}{ll l cc cc}
\toprule
\multirow{2}{*}{Dataset} &
\multirow{2}{*}{Task Name} &
\multirow{2}{*}{GNN Architecture} &
\multicolumn{2}{c}{Random} &
\multicolumn{2}{c}{\sysname{} (ours)} \\
\cmidrule(lr){4-5}\cmidrule(lr){6-7}
& & &
Test $\downarrow$ & Time/epoch $\downarrow$ &
Test $\downarrow$ & Time/epoch $\downarrow$ \\
\midrule
\multirow{3}{*}{rel-ratebeer} &
\multirow{3}{*}{user-count} &
HeteroGraphSAGE (3 hop) & 7.991 & 890 & \textbf{7.651} & \besttime{220} \\
& & HGT(3 hop)          & \textbf{12.431} & 3,740 & 14.447 & \besttime{736} \\
& & RelGT (3 hop, k=50)        & 8.387 & 2,086 & \textbf{7.960} & \besttime{900} \\
\midrule
\multirow{3}{*}{rel-trial} &
\multirow{3}{*}{study-adverse} &
HeteroGraphSAGE (3 hop) & \textbf{44.420} & 29 & 44.727 & \besttime{15} \\
& & HGT (3 hop)          & 57.379 & 80 & \textbf{57.118} & \besttime{45} \\
& & RelGT (3 hop, k=100)        & 46.012 & 388 & \textbf{44.984}  & \besttime{225}  \\
\midrule
\multirow{3}{*}{rel-trial} &
\multirow{3}{*}{site-success} &
HeteroGraphSAGE (3 hop) & \textbf{0.391} & 105 & 0.437 & \besttime{49} \\
& & HGT (3 hop)          & \textbf{0.462} & 410 & \textbf{0.462} & \besttime{124} \\
& & RelGT (3 hop, k=100)        & 0.394 & 1,720 & \textbf{0.369} & \besttime{744} \\
\midrule
\multirow{3}{*}{rel-f1} &
\multirow{3}{*}{driver-position} &
HeteroGraphSAGE (3 hop) & 4.973 & 14 & \textbf{4.317} & \besttime{4} \\
& & HGT (3 hop)           & 4.641 & 65 & \textbf{4.103} & \besttime{4} \\
& & RelGT (3 hop, k=100)        & 4.238 & 410 & \textbf{4.110}  & \besttime{84} \\
\midrule
\multirow{3}{*}{rel-stack} &
\multirow{3}{*}{post-votes} &
HeteroGraphSAGE (3 hop) & \textbf{0.065} & 1,084 & \textbf{0.065} & \besttime{442} \\
& & HGT (3 hop)           & \textbf{0.068} & 1,205 & \textbf{0.068} & \besttime{506} \\
& & RelGT (3 hop, k=50)        & \textbf{0.068} & 9,233 & \textbf{0.068} & \besttime{4,824} \\
\midrule
\multirow{3}{*}{rel-avito} &
\multirow{3}{*}{ad-ctr} &
HeteroGraphSAGE (3 hop) & \textbf{0.042} & 8 & \textbf{0.042} & \besttime{3} \\
& & HGT (3 hop)           & \textbf{0.043} & 8 & 0.044 & \besttime{4} \\
& & RelGT (3 hop, k=100)         & 0.039 & 152 & \textbf{0.038} & \besttime{104}  \\
\bottomrule
\end{tabular}
\caption{Node regression performance. Evaluation metric is MAE (lower is better). Epoch time is in seconds (lower is better). The fastest time/epoch is highlighted in green, and the best test metrics are in bold.}
\label{tab:node_regression_performance}
\vspace*{-16pt}
\end{table*}

Tables~\ref{tab:node_classification_performance} and~\ref{tab:node_regression_performance} present the main experimental results of this paper. Table~\ref{tab:node_classification_performance} reports test accuracy and per-epoch training time for the RelBench node classification tasks across the five databases in our evaluation, covering all three GNN backbones paired with the three sampling strategies. Table~\ref{tab:node_regression_performance} reports the same metrics for node regression tasks, but includes only random sampling and \sysname{}, \textbf{since MPS-GNN does not support node regression, a significant limitation relative to \sysname{}}.

\vspace{4pt}
\noindent
\textbf{Training time.}
Since the primary goal of \sysname{} is to reduce training time, we first examine per-epoch time. Both MPS-GNN and \sysname{} reduce training time compared to random neighborhood sampling, 
since they sample smaller subgraphs. However, \sysname{} delivers substantially larger gains on all tasks, with the exception of a few cases on \texttt{rel-stack} where the metapaths selected by MPS-GNN results in smaller subgraphs. Thus, we conclude that \textbf{\sysname{} achieves the fastest per-epoch time, yielding substantial savings over both MPS-GNN and random neighborhood sampling.} The speedup from \sysname{} can exceed 10$\times$,  for example, on the \texttt{beer-churn} task with the powerful RelGT backbone. This speedup reduces training time from days to hours.

\vspace{4pt}
\noindent
\textbf{Test accuracy.}
The training-time gains from \sysname{} and MPS-GNN do not come at the cost of accuracy. In fact, these sampling strategies often \emph{improve} accuracy, because the pruned metapaths are typically uninformative, so removing them from the training data yields a more accurate GNN. This effect is more pronounced for RelGT than for HeteroGraphSAGE and HGT. HeteroGraphSAGE and HGT are weaker GNN architectures and cannot fully exploit the improved training data produced by metapath selection. RelGT, in contrast, is a powerful GNN architecture designed specifically for RDL, and is therefore better positioned to leverage higher-quality training data. Specifically, RelGT is a transformer model trained on a fixed-size context window populated with tokens from sampled subgraphs. Metapath selection fills this window with higher-quality tokens. A clear takeaway from Tables~\ref{tab:node_classification_performance} and~\ref{tab:node_regression_performance} is that \textbf{the powerful RelGT backbone combined with \sysname{} metapath selection consistently achieves the highest accuracy.}

\begin{table}[t]
\centering
\small
\setlength{\tabcolsep}{2.5pt}
\renewcommand{\arraystretch}{1.15}
\resizebox{\columnwidth}{!}{%
\begin{tabular}{@{}llcccccc@{}}
\toprule
\multirow{2}{*}{Dataset} &
\multirow{2}{*}{Task Name} &
\multirow{2}{*}{\shortstack[c]{Train\\Samples}} &
\multirow{2}{*}{\shortstack[c]{Batch\\Count}} &
\multirow{2}{*}{\shortstack[c]{SQL\\Workers}} &
\multirow{2}{*}{\shortstack[c]{SQL\\Time}} &
\multicolumn{2}{c}{Total Time} \\
\cmidrule(lr){7-8}
& & & & & & Seconds & Epochs \\
\midrule
\multirow{3}{*}{rel-ratebeer}
& user-churn        & 300,000 & 32 & 4 & 638 & 649 & 0.07 \\
& beer-churn        & 300,000 & 32 & 4 & 2318 & 2340 & 0.13 \\
& user-count        & 300,000 & 32 & 4 & 652 & 669 & 0.32 \\
\midrule
\multirow{3}{*}{rel-trial}
& study-outcome     & 11,994  & 8  & 8 & 5    & 10    & 0.07 \\
& study-adverse     & 43,335  & 8  & 8 & 11    & 17    & 0.04 \\
& site-success      & 151,407 & 8  & 8 & 10    & 16    & 0.01 \\
\midrule
\multirow{3}{*}{rel-f1}
& driver-dnf        & 11,411  & 8  & 8 & 6    & 11    & 0.02 \\
& driver-top3       & 1,353   & 8  & 8 & 3    & 8    & 0.06 \\
& driver-position   & 7,453   & 8  & 8 & 5    & 11    & 0.03 \\
\midrule
\multirow{3}{*}{rel-stack}
& user-badge        & 363,048 & 32 & 8 & 31   & 55   & 0.02 \\
& user-engagement   & 168,020 & 32 & 8 & 24   & 35   & 0.01 \\
& post-votes        & 407,416 & 32 & 8 & 310   & 345   & 0.04 \\
\midrule
\multirow{3}{*}{rel-avito}
& user-visits       & 86,619  & 16 & 4 & 304   & 312   & 0.34 \\
& user-clicks       & 59,454  & 16 & 4 & 267  & 274   & 0.37 \\
& ad-ctr            & 5,100   & 8  & 8 & 54   & 59   & 0.39 \\
\bottomrule
\end{tabular}%
}
\caption{Time for \sysname{} to complete preprocessing and generate the required sampling configuration (hops = 3 for all tasks). SQL Time is in seconds. Total Time is reported in seconds, and as a fraction of the epoch time for RelGT with random neighborhood sampling for the same task.}
\label{tab:metasieve_preprocessing_durations}
\vspace*{-32pt}
\end{table}

\begin{table*}[t]
\centering
\small
\setlength{\tabcolsep}{4pt}
\renewcommand{\arraystretch}{1.15}
\begin{tabular}{@{}llcccccccc@{}}
\toprule
\multirow[b]{3}{*}{Dataset} &
\multirow[b]{3}{*}{Task Name} &
\multirow[b]{3}{*}{\shortstack[c]{Train\\Samples}} &
\multirow[b]{3}{*}{\shortstack[c]{Val\\Samples}} &
\multicolumn{3}{c}{Config ($k=3$, beam\_width=6)} &
\multicolumn{3}{c}{Config ($k=6$, beam\_width=15)} \\
\cmidrule(lr){5-7}\cmidrule(lr){8-10}
& & & &
\multicolumn{2}{c}{Total Time} &
\multirow[b]{2}{*}{\shortstack[c]{Paths\\Evaluated}} &
\multicolumn{2}{c}{Total Time} &
\multirow[b]{2}{*}{\shortstack[c]{Paths\\Evaluated}} \\
\cmidrule(lr){5-6}\cmidrule(lr){8-9}
& & & & Seconds & Epochs & & Seconds & Epochs & \\
\midrule
\multirow{2}{*}{rel-ratebeer}
& user-churn       & 500 & 120 & 3,532  & 0.27 & 22 & 3,790  & 0.39 & 30 \\
& beer-churn       & 500 & 120 & 7,922  & 0.45 & 26 & 11,257 & 0.64 & 51 \\
\midrule
rel-trial
& study-outcome    & 720 & 192 & 6,573  & 45.02 & 19 & 10,126 & 69.36 & 56 \\
\midrule
\multirow{2}{*}{rel-f1}
& driver-dnf       & 720 & 192 & 9,268  & 14.85 & 29 & 11,606 & 18.60 & 33 \\
& driver-top3      & 500 & 130 & 6,405  & 45.43 & 29 & 8,644  & 61.30 & 33 \\
\midrule
\multirow{2}{*}{rel-stack}
& user-badge       & 720 & 192 & 6,884  & 1.81 & 24 & 7,077  & 1.86 & 72 \\
& user-engagement  & 720 & 192 & 10,092 & 2.18 & 24 & 11,732 & 2.53 & 79 \\
\midrule
\multirow{2}{*}{rel-avito}
& user-visits      & 700 & 180 & 2,488  & 2.69 & 25 & 2,511  & 2.72 & 31 \\
& user-clicks      & 720 & 192 & 3,516  & 4.74 & 26 & 4,171  & 5.62 & 33 \\
\bottomrule
\end{tabular}
\caption{MPS-GNN search through hop 3 under the default ($k=3$, beam\_width=6) configuration and an alternative ($k=6$, beam\_width=15) configuration that produces more than 30 paths. Total Time is reported in seconds and as a multiple of the RelGT epoch time with random sampling. Paths Evaluated reports the number of metapaths evaluated during search.}
\label{tab:mps_search_durations}
\vspace*{-12pt}
\end{table*}

\begin{table*}[t]
\centering
\small
\setlength{\tabcolsep}{3.5pt}
\renewcommand{\arraystretch}{1.15}
\begin{tabular}{ll l cc cc cc}
\toprule
\multirow{2}{*}{Dataset} &
\multirow{2}{*}{Task Name} &
\multirow{2}{*}{GNN Architecture} &
\multicolumn{2}{c}{MPS-GNN (top 20)} &
\multicolumn{2}{c}{MPS-GNN (top 30)} &
\multicolumn{2}{c}{\sysname{}} \\
\cmidrule(lr){4-5}\cmidrule(lr){6-7}\cmidrule(lr){8-9}
& & &
Test $\uparrow$ & Time/epoch $\downarrow$ &
Test $\uparrow$ & Time/epoch $\downarrow$ &
Test $\uparrow$ & Time/epoch $\downarrow$ \\
\midrule
rel-ratebeer & user-churn
& RelGT (3 hop, k=50)
& 0.942 & 2,443
& 0.933 & 6,132
& \textbf{0.953} & \besttime{1,253} \\
\midrule
rel-ratebeer & beer-churn
& RelGT (3 hop, k=50)
& 0.780 & 11,102
& 0.793 & 14,070
& \textbf{0.799} & \besttime{1,494} \\
\midrule
rel-trial & study-outcome
& RelGT (3 hop, k=100)
& 0.676 & 119
& \textbf{0.705} & 137
& 0.702 & \besttime{97} \\
\midrule
rel-f1 & driver-dnf
& RelGT (3 hop, k=100)
& 0.681 & 331
& \textbf{0.774} & 528
& 0.736 & \besttime{115} \\
\midrule
rel-f1 & driver-top3
& RelGT (3 hop, k=100)
& 0.750 & 123
& 0.744 & 124
& \textbf{0.819} & \besttime{32} \\
\midrule
rel-stack & user-badge
& RelGT (3 hop, k=50)
& 0.877 & \besttime{2,720}
& \textbf{0.880} & 3,619
& 0.877 & 3,063 \\
\midrule
rel-stack & user-engagement
& RelGT (3 hop, k=50)
& 0.901 & \besttime{3,368}
& 0.904 & 4,423
& \textbf{0.905} & 3,697 \\
\midrule
rel-avito & user-visits
& RelGT (3 hop, k=100)
& 0.653 & 663
& 0.655 & 868
& \textbf{0.661} & \besttime{427} \\
\midrule
rel-avito & user-clicks
& RelGT (3 hop, k=50)
& 0.640 & 565
& 0.663 & 791
& \textbf{0.670} & \besttime{325} \\
\bottomrule
\end{tabular}
\caption{Node classification performance when MPS-GNN retains the top 20 and top 30 paths compared with \sysname{}. Evaluation metric is AUCROC (higher is better). Epoch time is in seconds (lower is better). The fastest time/epoch is highlighted in green, and the best test metrics are in bold.}
\label{tab:node_classification_performance_ablation_mpsgnn}
\vspace*{-16pt}
\end{table*}

\begin{table}[t]
\centering
\small
\setlength{\tabcolsep}{3.5pt}
\renewcommand{\arraystretch}{1.15}
\begin{tabular}{@{}llcccc@{}}
\toprule
\multirow[b]{2}{*}{Dataset} &
\multirow[b]{2}{*}{Task Name} &
\multicolumn{2}{c}{\sysname{}} &
\multicolumn{2}{c}{Random Pruning} \\
\cmidrule(lr){3-4}\cmidrule(lr){5-6}
& & \shortstack{Test\\\mbox{}} & \shortstack{Time/\\epoch} & \shortstack{Test\\\mbox{}} & \shortstack{Time/\\epoch} \\
\midrule
\multirow{3}{*}{rel-trial}
& study-outcome (cls $\uparrow$)  & 0.702 & 97 & 0.675 & 162 \\
& study-adverse (reg $\downarrow$)  & 44.984 & 225 & 46.566 & 345 \\
& site-success (reg $\downarrow$)   & 0.369 & 744 & 0.447 & 1,527 \\
\midrule
\multirow{3}{*}{rel-f1}
& driver-dnf (cls $\uparrow$)     & 0.736 & 115 & 0.703  & 149 \\
& driver-top3 (cls $\uparrow$)    & 0.766 & 29 & 0.795 & 31 \\
& driver-position (reg $\downarrow$) & 4.110 & 84 & 4.584 & 98 \\
\midrule
\multirow{3}{*}{rel-avito}
& user-visits (cls $\uparrow$)    & 0.661 & 427 & 0.647 & 529 \\
& user-clicks (cls $\uparrow$)    & 0.670 & 325 & 0.643 & 379 \\
& ad-ctr (reg $\downarrow$)         & 0.038 & 104 & 0.040 & 110 \\
\bottomrule
\end{tabular}
\caption{\sysname{} vs.\ random pruning using the same number of pruning rules starting from hop 2. All experiments use the RelGT backbone. For classification tasks (cls $\uparrow$), the test metric is AUCROC. For regression tasks (reg $\downarrow$), the test metric is MAE. Time/epoch is in seconds.}
\label{tab:random_metapath_dropping_ab}
\vspace*{-12pt}
\end{table}

\vspace*{-8pt}
\subsection{Preprocessing Time for Metapath Selection}
\label{sec:preprocessing}

Table~\ref{tab:metasieve_preprocessing_durations} reports the preprocessing time of \sysname{}. For each task, the table lists the number of training samples (seed nodes) and the number of batches $B$ into which they are divided. The SQL queries are parallelized into worker threads, and the number of workers is also listed. The next two columns give the time spent in SQL processing and the total preprocessing. SQL processing dominates, and the remaining steps (scoring metapaths, GMM clustering, and generating rules) take only 5--36 seconds. To place these numbers in context, the final column shows the preprocessing time divided by the average epoch time for training the RelGT backbone with random neighborhood sampling on the same task. We use RelGT because it is the strongest backbone in our experiments, and random neighborhood sampling because it is the baseline that \sysname{} is designed to improve upon. The results in this last column are striking: \textbf{the substantial gains in training time and test accuracy delivered by \sysname{} are obtained at a preprocessing cost of only a fraction of one epoch.}

Table~\ref{tab:mps_search_durations} reports the corresponding preprocessing time for MPS-GNN, restricted to classification tasks since MPS-GNN does not support regression. The first two columns list the number of training and validation samples used. We deliberately keep these sample sizes small as MPS-GNN trains a separate GNN at every decision step, and reducing the sample size is necessary to keep its running time tractable. The next two columns report the total preprocessing time, both in seconds and as a multiple of the average RelGT epoch time with random sampling on the same task. The remaining columns are discussed in the next section. Despite using far fewer samples than \sysname{}, MPS-GNN is substantially slower, taking up to 45 epochs. We train the GNN on the selected metapaths for at most 30 epochs, so a metapath selection procedure that can itself exceed this budget offers little practical value. In short, \textbf{MPS-GNN is too slow to be a viable choice in our setting}.

\vspace*{-4pt}
\subsection{Retaining More Metapaths in MPS-GNN}
\label{sec:sensitivity_mpsgnn}

To make sure that we are fair to MPS-GNN  in our experiments, we evaluate whether our choice of retaining only the top 20 metapaths it selects puts it at a disadvantage. In this section, we study the performance of MPS-GNN if we increase the number of evaluated metapaths and retain the top 30.

MPS-GNN does not provide direct control over the number of metapaths evaluated because this number depends on the available schema traversals and the candidates retained during the search at each hop. However, increasing the search specific parameters $k$ and \texttt{beam\_width} broadens the search and evaluates more metapaths. Our default setting uses $k=3$ and \texttt{beam\_width}$=6$.  We increase these values to $k=6$ and \texttt{beam\_width}$=15$ to increase the likelihood that at least 30 metapaths are evaluated. We retain the 30 highest scoring metapaths from the latter setting.

First, we study how this broader exploration affects preprocessing time. Table~\ref{tab:mps_search_durations} shows both the default search setting in which we retain 20 metapaths and the broader setting in which we retain 30 metapaths. For each setting, the table shows the preprocessing time in seconds and as a multiple of the RelGT epoch time. The table also shows the number of metapaths evaluated by MPS-GNN in both settings, confirming that this number is near or above 20 in the first setting and always equal or above 30 in the second. Thus, the second setting has more than 30 metapaths that are evaluated and scored by MPS-GNN, from which we can meaningfully select the top 30. However, the broader search for metapaths further increases MPS-GNN's already high preprocessing time, sometimes substantially (e.g., \texttt{rel-trial}). As expected, exploring and retaining more metapaths puts MPS-GNN at an even higher disadvantage to \sysname{} in terms of preprocessing cost.

The next question we ask is whether this increased preprocessing cost improves training. Table~\ref{tab:node_classification_performance_ablation_mpsgnn} shows the Test AUCROC and Time/Epoch for all classification tasks using the RelGT backbone. We compare MPS-GNN when retaining both top 20 and 30 metapaths, and \sysname{}.
Retaining 30 metapaths with MPS-GNN instead of 20 leads to larger sampled subgraphs, which increases time per epoch, sometimes by a large factor as in the \texttt{user-churn} task. This increase in training time is expected. The question is whether it leads to a more accurate model. Here, the answer is mixed. 
On one task, \texttt{driver-dnf}, retaining 30 metapaths substantially increases the test accuracy of MPS-GNN, making it the most accurate strategy by a good margin, albeit the most expensive.
On every other task, retaining 30 metapaths yields only a moderate gain in accuracy for MPS-GNN, or even a degradation (\texttt{user-churn} and \texttt{driver-top3}).
Thus, our default of retaining the top 20 metapaths selected by MPS-GNN does not place it at a disadvantage. On the contrary, in our view, it strikes a good balance between preprocessing and training time on one hand, and model accuracy on the other.

\begin{table*}[t]
\centering
\small
\setlength{\tabcolsep}{3.5pt}
\renewcommand{\arraystretch}{1.15}
\resizebox{\textwidth}{!}{%
\begin{tabular}{ll cc cc cc cc}
\toprule
\multirow[b]{2}{*}{Dataset} &
\multirow[b]{2}{*}{Task Name} &
\multicolumn{2}{c}{\sysname{} Formula} &
\multicolumn{2}{c}{Only Count} &
\multicolumn{2}{c}{Only Rate} &
\multicolumn{2}{c}{Only Coverage} \\
\cmidrule(lr){3-4}
\cmidrule(lr){5-6}
\cmidrule(lr){7-8}
\cmidrule(lr){9-10}
& & Test & Time/epoch & Test & Time/epoch & Test & Time/epoch & Test & Time/epoch \\
\midrule
\multirow{3}{*}{rel-trial}
& study-outcome (cls $\uparrow$) & 0.702 & 97 & 0.678 & 94 & 0.686 & 95 & 0.675 & 97 \\
& study-adverse (reg $\downarrow$) & 44.984 & 225 & 45.464 & 216 & 45.759 & 224 & 45.967 & 218 \\
& site-success (reg $\downarrow$) & 0.369 & 744 & 0.386 & 663 & 0.374 & 679 & 0.440 & 1,386 \\
\midrule
\multirow{3}{*}{rel-f1}
& driver-dnf (cls $\uparrow$)     & 0.736 & 115 & 0.726 & 182 & 0.706 & 109 & 0.699 & 99 \\
& driver-top3 (cls $\uparrow$)    & 0.766 & 29 & 0.675 & 45 & 0.800 & 39 & 0.728 & 29 \\
& driver-position (reg $\downarrow$) & 4.110 & 84 & 3.974 & 84 & 3.985 & 83 & 3.958 & 78 \\
\midrule
\multirow{3}{*}{rel-avito}
& user-visits (cls $\uparrow$) & 0.661 & 427 & 0.654 & 490 & 0.659 & 477 & 0.653 & 824 \\
& user-clicks (cls $\uparrow$) & 0.670 & 325 & 0.649 & 371 & 0.651 & 372 & 0.615 & 619 \\
& ad-ctr (reg $\downarrow$)    & 0.038 & 104 & 0.039 & 111 & 0.038 & 112 & 0.039 & 102 \\
\bottomrule
\end{tabular}%
}
\caption{Effect of different components of the formula for $Q$. All experiments use the RelGT backbone. For classification tasks (cls $\uparrow$), the test metric is AUCROC. For regression tasks (reg $\downarrow$), the test metric is MAE. Time/epoch is in seconds.}
\label{tab:ablation_empty_template}
\vspace*{-12pt}
\end{table*}

\subsection{Ablations on the Scoring Function $Q$}
\label{sec:ablation}

Next, we turn our attention to the internal hyperparameters of \sysname{}. In this section, we present ablation studies on the scoring function $Q(\pi_{1:h},e)$. In the next two sections, we examine hyperparameter $\delta$ and the heuristic used to generate sampling rules. All our ablation studies use the RelGT backbone on the same three databases:
\texttt{rel-trial},
\texttt{rel-f1},
and
\texttt{rel-avito}.

The first ablation study on $Q(\pi_{1:h},e)$ asks whether this scoring function is necessary at all: \sysname{} uses $Q$ to decide which metapaths to prune, and we ask whether pruning the same number of metapaths uniformly at random would achieve comparable results. For each task, let $n$ denote the number of pruning rules produced by \sysname{}. We randomly generate exactly $n$ rules by choosing $n$ schema-valid (prefix, candidate extension) pairs from the set produced by schema-guided metapath enumeration (Section~\ref{sec:enumeration}). As in \sysname{}, we begin pruning only from hop~2 onward, so random pruning has the same opportunity to remove extensions at the same stages of the search space. Table~\ref{tab:random_metapath_dropping_ab} reports the results. Random pruning is consistently inferior to \sysname{} in both accuracy and training time, confirming the need for \sysname{}'s scoring.

The second ablation study asks whether all components of the formula for $Q$ are necessary. Recall that
\vspace*{-4pt}
\[
Q(\pi_{1:h},e)
=
\max\!\left(p_{\mathrm{log\text{-}count}}(\pi_{1:h},e),\, p_{\mathrm{log\text{-}rate}}(\pi_{1:h},e)\right)
+
p_{\mathrm{cov}}(\pi_{1:h},e).
\]
\vspace*{-2pt}
We compare the full scoring function against three simpler variants that score extensions using only $p_{\mathrm{log\text{-}count}}$, only $p_{\mathrm{log\text{-}rate}}$, or only $p_{\mathrm{cov}}$. Table~\ref{tab:ablation_empty_template} reports the results. While individual components are occasionally competitive, the full formula consistently produces the best accuracy, confirming the need for all components.

\subsection{Tuning Hyperparameter $\delta$}
\label{sec:delta-tuning}

Recall from Section~\ref{sec:q_score} that the hyperparameter $\delta$ controls the LCB computation and hence the score $Q$. This affects the metapath extensions in the good vs.\ bad cluster produced by the GMM (Section~\ref{sec:thresholding}).
To choose $\delta$ for a task, we train the GNN by pruning metapaths using rules generated with $\delta \in \{0.1, 0.2, 0.3, 0.4\}$ and evaluate on the \emph{validation} split of the data.
Table~\ref{tab:delta_ablation_val} shows the results.

In some tasks, changing $\delta$ modifies the candidate scores of the metapaths and their ranking but leaves the GMM split unchanged. That is, the same extensions end up in the good and bad clusters, and the same extensions are therefore pruned.
This can be seen in Table~\ref{tab:delta_ablation_val} in the form of relatively unchanging time per epoch and validation accuracy, 
as observed for \texttt{study-adverse} and \texttt{user-clicks}.

For other tasks, increasing $\delta$ moves borderline extensions across the GMM boundary, thus changing the good and bad clusters. If the bad cluster shrinks, fewer metapaths are pruned and time per epoch increases, as seen for \texttt{site-success} at $\delta=0.40$. The opposite can also occur. If the bad cluster grows, more metapaths are pruned and time per epoch decreases, as illustrated by \texttt{driver-top3} going from $\delta=0.10$ to $\delta=0.20/0.30$. This effect is not monotonic because the hop wise percentile normalization and GMM clustering depend on the relative separation between candidate extensions rather than only on their absolute LCB values.

We select a task-specific $\delta$ using validation accuracy as the primary criterion and time per epoch as the secondary criterion when the validation results are close. 
Table~\ref{tab:delta_ablation_val} shows the selected $\delta$ for each task.
In most cases, we select the $\delta$ value with the best validation accuracy, but in some cases we accept a small degradation in accuracy when it provides a substantial reduction in training time. For example, \texttt{study-outcome} uses $\delta=0.40$ instead of $\delta=0.20$, while \texttt{user-visits} uses $\delta=0.10$ instead of $\delta=0.30$. When all tested $\delta$ values produce the same pruning rules, the task is insensitive to $\delta$, so we choose $\delta=0.40$.

Overall, $\delta$ affects pruning and performance for some tasks. However, \sysname{} is not overly sensitive to the choice of $\delta$ to the point of requiring unreasonable effort in tuning this hyperparameter or breaking down if a wrong value is chosen. Obtaining the best performance requires tuning $\delta$, but it would not be unreasonable to use a default of, say, $\delta = 0.2$ for all tasks.

\begin{table*}[t]
\centering
\small
\setlength{\tabcolsep}{5pt}
\renewcommand{\arraystretch}{1.15}
\resizebox{\textwidth}{!}{%
\begin{tabular}{ll cc cc cc cc c}
\toprule
\multirow[b]{2}{*}{Dataset} &
\multirow[b]{2}{*}{Task Name} &
\multicolumn{2}{c}{$\delta = 0.10$} &
\multicolumn{2}{c}{$\delta = 0.20$} &
\multicolumn{2}{c}{$\delta = 0.30$} &
\multicolumn{2}{c}{$\delta = 0.40$} &
\multirow{2}{*}{\shortstack{Selected\\$\delta$}} \\
\cmidrule(lr){3-4}
\cmidrule(lr){5-6}
\cmidrule(lr){7-8}
\cmidrule(lr){9-10}
& & Val & Time/epoch & Val & Time/epoch & Val & Time/epoch & Val & Time/epoch & \\
\midrule
\multirow{3}{*}{rel-trial}
& study-outcome (cls $\uparrow$) & 0.670 & 109 & 0.670 & 108 & 0.647 & 97 & 0.665 & 97 & 0.40 \\
& study-adverse (reg $\downarrow$) & 47.486 & 225 & 47.878 & 222 & 47.685 & 225 & 47.534 & 225 & 0.40 \\
& site-success (reg $\downarrow$) & 0.379 & 743 & 0.369 & 744 & 0.382 & 732 & 0.380 & 1,547 & 0.20 \\
\midrule
\multirow{3}{*}{rel-f1}
& driver-dnf (cls $\uparrow$) & 0.769 & 100 & 0.750 & 111 & 0.763 & 116 & 0.782 & 115 & 0.40 \\
& driver-top3 (cls $\uparrow$) & 0.741 & 55 & 0.856 & 30 & 0.881 & 29 & 0.820 & 32 & 0.30 \\
& driver-position (reg $\downarrow$) & 3.094 & 84 & 3.349 & 86 & 3.580 & 85 & 3.222 & 104 & 0.10 \\
\midrule
\multirow{3}{*}{rel-avito}
& user-visits (cls $\uparrow$) & 0.687 & 427 & 0.689 & 705 & 0.691 & 676 & 0.690 & 675 & 0.10 \\
& user-clicks (cls $\uparrow$) & 0.634 & 325 & 0.639 & 326 & 0.637 & 323 & 0.637 & 325 & 0.40 \\
& ad-ctr (reg $\downarrow$) & 0.034 & 106 & 0.035 & 105 & 0.033 & 104 & 0.034 & 104 & 0.40 \\
\midrule
\end{tabular}%
}
\caption{Validation performance for different tasks across a range of values of hyperparameter $\delta$. All experiments use the RelGT backbone. For classification tasks (cls $\uparrow$), the metric is AUCROC. For regression tasks (reg $\downarrow$), the metric is MAE.}
\vspace*{-12pt}
\label{tab:delta_ablation_val}
\end{table*}

\begin{table*}[t]
\centering
\small
\setlength{\tabcolsep}{3.5pt}
\renewcommand{\arraystretch}{1.15}
\begin{tabular}{lll c ccc ccc ccc}
\toprule
\multirow[b]{2}{*}{Dataset} &
\multirow[b]{2}{*}{Task Name} &
\multirow[b]{2}{*}{GNN Architecture} &
\multirow[b]{2}{*}{Hop} &
\multicolumn{3}{c}{Default Heuristic} &
\multicolumn{3}{c}{Majority Heuristic} &
\multicolumn{3}{c}{Unanimous Heuristic} \\
\cmidrule(lr){5-7}
\cmidrule(lr){8-10}
\cmidrule(lr){11-13}
& & & &
Test & Time/epoch $\downarrow$ & \shortstack{No. of\\Rules} &
Test & Time/epoch $\downarrow$ & \shortstack{No. of\\Rules} &
Test & Time/epoch $\downarrow$ & \shortstack{No. of\\Rules} \\
\midrule

rel-f1 &
driver-dnf (cls $\uparrow$) &
RelGT (k=100) &
3 &
0.736 & 115 & 10 &
0.701 & 189 & 7 &
0.701 & 189 & 7 \\

\midrule

rel-f1 &
driver-position (reg $\downarrow$) &
RelGT (k=100) &
3 &
4.110 & 84 & 11 &
4.005 & 134 & 7 &
4.005 & 134 & 7 \\

\midrule

rel-stack &
user-badge (cls $\uparrow$) &
HeteroGraphSAGE &
4 &
0.889 & 185 & 129 &
0.890 & 207 & 122 &
0.890 & 236 & 58 \\

rel-stack &
user-badge (cls $\uparrow$) &
RelGT (k=50) &
3 &
0.877 & 3,063 & 23 &
0.877 & 3,063 & 23 &
0.901 & 3,508  & 1 \\

\midrule

rel-stack &
user-engagement (cls $\uparrow$) &
HeteroGraphSAGE &
4 &
0.906 & 209 & 172 &
0.907 & 214 & 166 &
0.907 & 217 & 151 \\

rel-stack &
user-engagement (cls $\uparrow$) &
RelGT (k=50) &
3 &
0.905 & 3,697 & 23 &
0.905 & 3,812 & 21 &
 0.904 & 4,636  & 0 \\

\midrule

rel-stack &
post-votes (reg $\downarrow$) &
RelGT (k=50) &
3 &
0.068 & 4,824 & 23 &
0.068 & 8,371 & 4 &
0.068 & 8,371 & 4 \\

\bottomrule
\end{tabular}
\caption{Performance under different heuristics for generating sampling rules. For classification tasks (cls $\uparrow$), the test metric is AUCROC. For regression tasks (reg $\downarrow$), the test metric is MAE. Time/epoch is in seconds. The number of pruning rules decreases as the heuristic becomes less aggressive in recommending pruning.}
\label{tab:sensitivity_analysis}
\vspace*{-16pt}
\end{table*}

\vspace*{-12pt}
\subsection{Prefix-aware Sampling Rules}
\label{sec:prefix-agnostic-rule-approximation}

As discussed in Section~\ref{sec:rule_generation}, a candidate extension $e$ in a prefix-aware rule at hop $h$ for prefix $\pi_{1:h}$ may also be reachable through other prefixes ending at the same terminal table. Since PyG does not track the prefix used to reach a node, the rule for $e$ at hop $h$ is applied to all such prefixes. We examine how often this causes \emph{overpruning}. Specifically, for each hop and terminal table appearing in the pruning rules, we check all schema-valid prefixes that can reach that terminal table and compare them with the prefixes present in the rules. If all such prefixes are present, applying the prefix-agnostic rule does not introduce additional pruning. If some prefixes are missing, the rule may prune $e$ when the terminal table is reached through those missing prefixes, which we refer to as overpruning.

We start with the non-problematic cases. For all tasks on the \texttt{rel-trial} dataset up to three hops, each affected terminal table is reachable through only one prefix, so no conflict between prefixes is possible. Tasks on \texttt{rel-avito} and \texttt{rel-ratebeer} often have multiple prefixes reaching the same terminal table, but all prefixes are represented in the pruning rules, so they all agree on pruning the extension.
The same is true for the \texttt{rel-f1/driver-top3} task.
In all these cases, no overpruning is possible.

\begin{sloppypar}
We now turn to cases where overpruning occurs. In the \texttt{rel-f1/driver-dnf} and \texttt{rel-f1/driver-position} tasks, one of the two prefixes reaching the \texttt{constructors} table at hop 2 is missing from the pruning rules, so \texttt{constructors} rows reachable through this prefix are overpruned.
Overpruning is more common in the \texttt{rel-stack} dataset because its schema allows many metapaths to converge on the same terminal table. The number of prefix metapaths missing from the pruning rules is small at three hops but becomes larger at four hops as the number of valid metapaths grows. We observe this pattern across the \texttt{user-badge}, \texttt{user-engagement}, and \texttt{post-votes} tasks, showing that overpruning is more likely in highly connected schemas and at greater hop depths.
\end{sloppypar}

To measure the practical effect of overpruning, Table~\ref{tab:sensitivity_analysis} compares three heuristics for generating sampling rules. The first is the default, prefix-agnostic heuristic of \sysname{}, which prunes an extension if it appears in the pruning rules with any of the prefixes leading to it, regardless of how many prefixes are missing from the pruning rules. We contrast this with two less aggressive heuristics, which we call \emph{majority} and \emph{unanimous}. The majority heuristic retains a sampling rule generated by the default heuristic for a hop and terminal table only when the number of prefixes reaching this table at this hop in the \sysname{} pruning rules is greater than the number of prefixes missing from the pruning rules. That is, a majority of the prefix metapaths reaching an extension recommends pruning this extension. The unanimous heuristic retains a sampling rule generated by the default heuristic only when every prefix is present in the pruning rules. Table~\ref{tab:sensitivity_analysis} reports the number of sampling rules under each heuristic. Note that the rules under unanimous are a subset of the rules under majority, which in turn are a subset of the rules under default.

Table~\ref{tab:sensitivity_analysis} also shows the test accuracy and per-epoch time when using the different heuristics. The table shows results for two \texttt{rel-stack} tasks with 4 hops, since overpruning becomes a greater problem with higher hop counts. In contrast to the other rows, which use RelGT as a backbone, these two rows use HeteroGraphSAGE since training RelGT for 4 hops is prohibitively expensive. 
The results show that reducing overpruning by using less aggressive heuristics does not improve predictive accuracy. The majority and unanimous heuristics perform slightly better than the default heuristic in a few cases and worse in some, but the key takeaway is that differences in accuracy are minimal.
This suggests that the GNNs are robust to the structural information that the default heuristic loses. The clearer effect is on training cost. The less aggressive heuristics retain additional graph structure, producing larger sampled neighborhoods and slower epochs. Overall, the default, prefix-agnostic heuristic provides the strongest balance between predictive accuracy and training efficiency.

In summary, the prefix-agnostic heuristic used by \sysname{} to generate sampling rules introduces no overpruning in most task and hop settings. Overpruning is most noticeable in highly connected schemas, such as \texttt{rel-stack}, and at deeper hop counts. Even in these cases, retaining more structure through the majority or unanimous heuristics changes predictive performance only slightly while increasing training time. This shows that the \sysname{} heuristic is a practical choice.

\begin{table*}[t]
\centering
\small
\setlength{\tabcolsep}{3.5pt}
\renewcommand{\arraystretch}{1.15}
\begin{tabular}{@{}lcccccccc@{}}
\toprule
Dataset/Task Name & Workers & Batches & \# Cand.
  & \shortstack{Task Frontier\\Rows}
  & \shortstack{Peak Child\\Rows}
  & \shortstack{Concurrent Peak\\Child Rows}
  & \shortstack{Global Peak\\RSS (GB)}
  & \shortstack{Wall-Clock\\Time (s)} \\
\midrule
rel-ratebeer/beer-churn   & 4 & 32 & 65  & 124.13B & 5.28B   & 16.73B  & 1,040.22 & 1,436 \\
rel-ratebeer/user-churn   & 4 & 32 & 51  & 51.51B  & 3.59B   & 13.25B  & 630.45   & 471   \\
rel-avito/user-visits     & 4 & 16 & 23  & 25.67B  & 878.28M & 3.34B   & 280.34   & 155   \\
rel-avito/user-clicks     & 4 & 16 & 23  & 21.85B  & 758.26M & 2.92B   & 233.91   & 135   \\
rel-stack/post-votes      & 8 & 32 & 277 & 13.02B  & 153.23M & 581.94M & 147.65   & 168   \\
rel-stack/user-badge      & 8 & 32 & 90  & 252.45M & 7.38M   & 9.14M   & 12.77    & 31    \\
rel-stack/user-engagement & 8 & 32 & 90  & 217.78M & 7.31M   & 7.42M   & 11.22    & 30    \\
\bottomrule
\end{tabular}
\caption{Frontier materialization and memory pressure under multi-worker execution. Task Frontier Rows is the total materialized volume, Peak Child Rows is the largest single frontier, and Concurrent Peak Child Rows is the largest combined frontier size of concurrently active queries. Global Peak RSS is the maximum sampled process RSS, and Wall-Clock Time is the end-to-end runtime. M and B denote million and billion rows.}
\label{tab:operational_frontier_memory}
\vspace*{-12pt}
\end{table*}

\vspace*{-8pt}
\subsection{Scalability of SQL Operations}
\label{sec:scalability}
\sysname{} uses SQL operations (queries and create table statements) to create the data used in metapath selection, and in this section, we study the scalability of these operations as the database workload increases. \sysname{} performs two main SQL operations for each candidate extension: frontier materialization followed by aggregation on the frontier to compute per-seed log-count and log-rate statistics. 
We focus our scalability analysis on frontier materialization because this is where a candidate extension can substantially increase the number of intermediate rows. When an extension enters a dense or high-fanout region of the database, a relatively small parent frontier can expand into a very large child frontier, increasing memory pressure. In contrast, aggregation to compute statistics reduces an already materialized frontier; although it contributes to preprocessing time, it does not create the frontier expansion itself. Thus, \sysname{} must control frontier-induced memory pressure to scale reliably on large, highly connected databases.

As the database size grows, the time and memory required for frontier materialization both grow. Table~\ref{tab:metasieve_preprocessing_durations} shows that this time is small, a fraction of an epoch, even for large databases such as \texttt{rel-ratebeer} and \texttt{rel-avito}. In this section, we study the memory required for frontier materialization by profiling the memory growth of candidate-extension queries in different settings. This enables us to answer the following questions:
How much memory is required for frontier materialization?
Can the amount of memory be regulated?
How does regulating memory affect execution time?

We focus on the largest databases in Table~\ref{tab:metasieve_preprocessing_durations}: \texttt{rel-ratebeer}, \texttt{rel-stack}, and \texttt{rel-avito}. The remaining databases, \texttt{re-trial} and \texttt{rel-f1}, are small and do not create memory pressure.

\subsubsection{Memory Pressure of Frontier Materialization}

We first examine how the size of a single materialized child frontier affects memory pressure. Because each candidate extension $(\pi_{1:h},e)$ is evaluated independently over multiple seed batches, it can produce different frontier sizes across batches. For each candidate extension, we select the batch
\[
b^{*}(\pi_{1:h},e)
=
\arg\max_b
\left|
\mathcal{F}^{(b)}_{h+1}(\pi_{1:h}\circ e)
\right|,
\]
where $\mathcal{F}^{(b)}_{h+1}(\pi_{1:h}\circ e)$ denotes the frontier table materialized for batch $b$ and for the candidate extension $(\pi_{1:h},e)$. This captures the worst observed memory pressure for that extension.

We quantify the memory pressure created by frontier materialization in terms of the increase in the \emph{resident set size (RSS)} of the \sysname{} process (recall that DuckDB runs inside this process). For the frontier-materialization query of candidate extension $(\pi_{1:h},e)$ in its selected batch $b^{*}(\pi_{1:h},e)$, we measure the increase in resident set size as
\[
\Delta\mathrm{RSS}(\pi_{1:h},e)
=
\mathrm{RSS}_{\mathrm{peak}}
-
\mathrm{RSS}_{\mathrm{before}}
\]
where $\mathrm{RSS}_{\mathrm{before}}$ is the process RSS immediately before executing the query and $\mathrm{RSS}_{\mathrm{peak}}$ is the largest RSS observed during its execution. RSS is sampled every 50\,ms. We use one worker and one query execution thread so that frontier-materialization queries do not overlap, allowing the measured RSS increase to be attributed to a single candidate extension rather than concurrent SQL work.

\begin{figure}[t]
    \centering
    \includegraphics[width=\columnwidth]{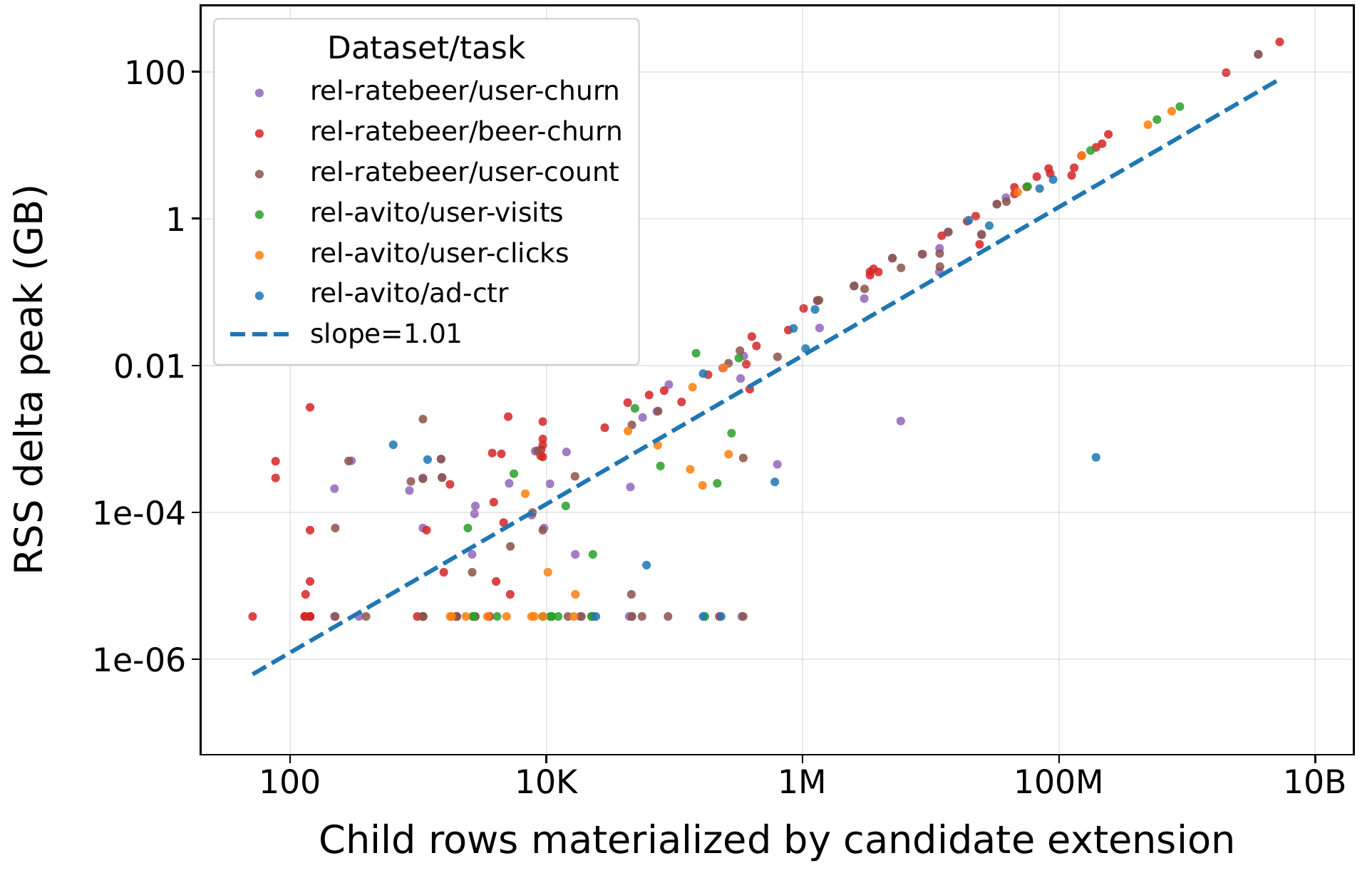}
    \caption{Memory pressure of materializing a single frontier. Each point represents a candidate extension for a task. The x-axis represents the largest frontier table generated for this candidate extension. The y-axis is the peak RSS increase for the SQL query that generated this largest frontier table. Both axes use a log scale.}
    \label{fig:frontier_memory_pressure}
    \vspace*{-12pt}
\end{figure}

Figure~\ref{fig:frontier_memory_pressure} shows the increase in RSS for six of the largest RelBench tasks. The frontiers vary widely in size, and the increase in RSS grows approximately linearly with the number of materialized child rows, particularly for the largest frontiers on the right side of the plot. 
The figure also shows the best-fit line for the points, which has a slope of $1.01$, a very slightly superlinear relation.
Such an almost-linear increase is good, and indicates that the memory for frontier materialization does not explode, even for the largest frontiers.

For many small-to-medium frontiers, 
the measured increase remains near $4$\,KB, forming the horizontal line of points near the bottom of the plot. 
The memory page size is $4$\,KB, and the horizontal line indicates that DuckDB often reuses memory already allocated by earlier queries, leaving no observable RSS growth beyond 1 page, the operating system's page-level measurement granularity. 
Once frontiers become sufficiently large, additional memory is required, and RSS grows in proportion to frontier size.

\subsubsection{Effect of Multi-worker Execution}
\label{sec:multiworker_memory_pressure}

The preceding experiment quantifies
the memory pressure of frontier materialization using one worker and one query execution thread. However, \sysname{} normally processes multiple seed batches concurrently, as seen in Table~\ref{tab:metasieve_preprocessing_durations}, using multiple workers and multiple execution threads per worker.
Consequently, several candidate-extension queries may materialize child frontiers concurrently, resulting in greater memory pressure than an isolated query. In this section, we quantify memory pressure under multi-worker execution.

Table~\ref{tab:operational_frontier_memory} reports three complementary measures of frontier materialization under multi-worker execution for seven of the largest tasks, sorted in descending order of these measures. 
Each row in the table reports the task name, number of workers used, number of batches, and the number of candidate extensions for this task.
Next, the table reports the three measures of frontier materialization.
\emph{Task Frontier Rows} is the total number of child-frontier rows materialized by the entire task. It captures the cumulative frontier-processing workload and helps explain the overall execution time, since tasks that materialize more rows require more total SQL processing. \emph{Peak Child Rows} is the largest child frontier produced by any single candidate-extension query in any batch. This metric isolates the individual query that creates the largest materialization bottleneck.

These two metrics do not capture the additional pressure created when several queries execute concurrently. We therefore measure \emph{Concurrent Peak Child Rows}. For each candidate-extension query, we record when it starts and ends, and the number of rows in its materialized child frontier. At every point during the run, we identify queries that are executing concurrently and sum their child-frontier sizes. The largest such sum over the complete run is the Concurrent Peak Child Rows. This metric captures the maximum 
amount of frontier materialization active simultaneously 
and therefore serves as a proxy for memory pressure under multi-worker execution.

The next column in the table, \emph{Global Peak RSS}, reports the corresponding memory consumption in GB. We sample the resident set size of the \sysname{} process during frontier generation every $0.5$ seconds. 
Global Peak RSS is the largest observed RSS value. Since all workers execute within the same process, this measurement captures their combined memory footprint. Note that this table and the remaining tables in the paper all show RSS not $\Delta$RSS. The final column shows end-to-end \emph{Wall-Clock Time}.

The results show that larger values of Concurrent Peak Child Rows correspond to larger Global Peak RSS values. 
That is, memory pressure in multi-worker execution is largely determined by the combined size of the child frontiers materialized simultaneously.
For example, \texttt{beer-churn} has the largest concurrent peak at $16.73$ billion rows and reaches the highest memory usage at $1{,}040.22$\,GB. In contrast, \texttt{user-badge} and \texttt{user-engagement} have concurrent peaks below $10$ million rows and remain near $12$\,GB.

\begin{table*}[t]
\centering
\small
\setlength{\tabcolsep}{4pt}
\renewcommand{\arraystretch}{1.15}
\begin{tabular}{@{}lccccc@{}}
\toprule
Dataset/Task Name & Workers & Batches
  & \shortstack{Concurrent Peak\\Child Rows}
  & \shortstack{Global Peak\\RSS (GB)}
  & \shortstack{Wall-Clock\\Time (s)} \\
\midrule
rel-ratebeer/beer-churn & 4 & 32 & 16.73B & 1,040.22 & 1,436 \\
rel-ratebeer/beer-churn & 3 & 32 & 13.87B & 918.35   & 1,803 \\
rel-ratebeer/beer-churn & 2 & 32 & 10.34B & 697.91   & 2,598 \\
\midrule
rel-ratebeer/user-churn & 6 & 32 & 20.04B & 974.21   & 390 \\
rel-ratebeer/user-churn & 5 & 32 & 16.61B & 801.63   & 420 \\
rel-ratebeer/user-churn & 4 & 32 & 13.25B & 630.45   & 471 \\
\bottomrule
\end{tabular}
\caption{Effect of reducing worker count on memory. Reducing the number of workers reduces memory pressure but increases wall-clock time. B denotes billion rows.}
\label{tab:ratebeer_worker_scaling}
\vspace*{-12pt}
\end{table*}

\begin{table*}[t]
\centering
\small
\setlength{\tabcolsep}{4pt}
\renewcommand{\arraystretch}{1.15}
\begin{tabular}{@{}lccccc@{}}
\toprule
Dataset/Task Name & Workers & Batches
  & \shortstack{Concurrent Peak\\Child Rows}
  & \shortstack{Global Peak\\RSS (GB)}
  & \shortstack{Wall-Clock\\Time (s)} \\
\midrule
rel-ratebeer/beer-churn & 4 & 32 & 16.73B & 1,040.22 & 1,436 \\
rel-ratebeer/beer-churn & 4 & 48 & 12.77B & 792.60   & 1,420\\
rel-ratebeer/beer-churn & 4 & 64 & 10.34B & 637.47   & 1,427 \\
\midrule
rel-ratebeer/user-churn & 6 & 32 & 20.04B & 974.21   & 390   \\
rel-ratebeer/user-churn & 6 & 48 & 13.25B & 613.54   & 397   \\
rel-ratebeer/user-churn & 6 & 64 & 10.17B & 451.60   & 375  \\
\bottomrule
\end{tabular}
\caption{Effect of increasing batch count on memory. More batches place fewer seeds in each batch, reducing memory pressure. B denotes billion rows.}
\label{tab:ratebeer_batch_scaling}
\vspace*{-12pt}
\end{table*}

\begin{table*}[t]
\centering
\small
\setlength{\tabcolsep}{4pt}
\renewcommand{\arraystretch}{1.15}
\begin{tabular}{@{}lcccccc@{}}
\toprule
Dataset/Task Name & Workers
  & \shortstack{Mem Limit\\/Worker}
  & \shortstack{Threads\\/Worker}
  & \shortstack{Peak Spill\\(GB)}
  & \shortstack{Global Peak\\RSS (GB)}
  & \shortstack{Wall-Clock\\Time (s)} \\
\midrule
rel-ratebeer/beer-churn & 4 & default & 8 & 10.64 & 1,040.22 & 1,436 \\
rel-ratebeer/beer-churn & 4 & 128\,GB & 8 & 375.25 & 520.33 & 1,335 \\
rel-ratebeer/beer-churn & 4 & 64\,GB  & 4 & 489.83 & 272.80 & 1,608 \\
rel-ratebeer/beer-churn & 4 & 32\,GB  & 2 & 504.13 & 142.15 & 2,253 \\
rel-ratebeer/beer-churn & 4 & 16\,GB  & 2 & 544.30 & 76.81 & 2,150 \\
\midrule
rel-ratebeer/user-churn & 6 & default & 8 &   0.00 & 974.21 &   390 \\
rel-ratebeer/user-churn & 6 & 96\,GB  & 8 & 301.00 & 577.74 &   312 \\
rel-ratebeer/user-churn & 6 & 48\,GB  & 4 & 463.67 & 304.18 &   390 \\
rel-ratebeer/user-churn & 6 & 24\,GB  & 2 & 534.75 & 160.40 &   565 \\
rel-ratebeer/user-churn & 6 & 12\,GB  & 2 & 574.84 &  84.90 &   547 \\
\bottomrule
\end{tabular}
\caption{Effect of per-worker memory and thread limits on memory. A lower memory limit forces DuckDB to spill intermediate data to disk, substantially reducing Global Peak RSS while increasing disk usage and Wall-Clock Time. Peak Spill is the maximum sampled size of the temporary spill files.}
\label{tab:ratebeer_spilling}
\vspace*{-12pt}
\end{table*}

\vspace*{-4pt}
\subsubsection{Regulating Memory Pressure}
\label{sec:regulating_memory_pressure}

Up to this point, we were not restricting the memory available to \sysname{}.
We never experienced out-of-memory errors on our machine, which has 2\,TB of RAM --- not small, but very typical for modern servers. Nevertheless, it is important to have mechanisms that regulate the memory used by \sysname{}, to accommodate larger databases or smaller servers. In this section, we study these mechanisms.

Regulating the memory consumed by a database workload is a standard exercise that is well-supported by modern database systems. \sysname{} provides three practical control mechanisms for regulating memory pressure: First, we could reduce the number of concurrent workers, thereby directly limiting the concurrent pressure on memory. Second, we could increase the number of seed-node batches, which makes each batch smaller. Third, we could reduce the memory and worker threads given to DuckDB, relying on its ability to spill to disk as needed. These control mechanisms expose different trade-offs between memory consumption, execution time, and disk spilling.

In this experiment, we focus on the two largest tasks: \texttt{beer-churn} and \texttt{user-churn} on the \texttt{rel-ratebeer} dataset.
Our default setting is to use 
4 workers for \texttt{beer-churn} and 6 workers for \texttt{user-churn}, since \texttt{beer-churn} is a larger task. 
Recall that each worker gets 8 query execution threads for DuckDB by default, and a memory limit of \(M_{\mathrm{worker}} = 0.65M_{\mathrm{available}}/W\), where \(M_{\mathrm{available}}\) is the available system memory and \(W\) is the number of workers. 
For both tasks, our default is to use 32 seed-node batches.
This default configuration is designed to maximize execution speed at the expense of high memory usage. Our goal is to study the effectiveness of the three memory control mechanisms described above in reducing memory consumption relative to the default setting, and to measure their impact on execution time.

Table~\ref{tab:ratebeer_worker_scaling} shows how some of the metrics introduced in Table~\ref{tab:operational_frontier_memory} vary as we reduce the number of workers.
Reducing the number of workers limits the number of frontier materialization queries that can execute simultaneously, which reduces memory pressure. However, fewer workers means fewer resources dedicated to frontier materialization, so wall-clock time increases, but the table shows no unmanageable jumps in wall-clock time as the number of workers decreases.
Nevertheless, we see next that the other two mechanisms are more effective than reducing the number of workers.

Table~\ref{tab:ratebeer_batch_scaling} shows the effect of increasing the number of seed-node batches.
The total number of seed nodes remains unchanged, so each batch contains fewer seeds. Therefore, candidate-extension queries materialize smaller frontiers, reducing both Concurrent Peak Child Rows and Global Peak RSS. What is interesting is that increasing the number of batches leaves the overall runtime nearly unchanged. The reasons are as follows: First, increasing the number of batches does not reduce the resources dedicated to frontier materialization. Second, more batches create more SQL queries, but each query has a smaller working set. In the tested range (32--64 batches), the benefit from avoiding large, memory-heavy joins outweighs the fixed overhead of issuing more SQL statements. Third, one extremely large batch no longer dominates the wall-clock time, and from a large pool of independent units of work (more queries due to more batches), new queries can be selected for execution.

The final mechanism for regulating \sysname{} memory is to reduce each worker's memory limit and thread allocation while keeping the worker and batch counts fixed.
Table~\ref{tab:ratebeer_spilling} shows the effect of reducing the memory limit per worker. As the memory limit is reduced, the number of threads per worker must also be reduced to reduce the parallel work within each worker and ensure that each thread has enough memory for effective execution. Lowering the memory limit forces DuckDB to spill intermediate data to disk rather than retaining it in memory, which is the standard way DuckDB and other database systems handle memory pressure.
The table reports Peak Spill, the largest total spill file size observed across all workers, measured at $0.5$-second intervals. The default setting results in minimal disk spilling but high memory consumption. Reducing the memory limit of each worker substantially reduces memory consumption as measured by Global Peak RSS, but increases disk usage and often execution time because spilled intermediate data must be written to and read from disk. Reducing the memory limit is the most direct way to run \sysname{} within a small memory budget, relying on the memory management capabilities of DuckDB (or any other DBMS). The table shows that the memory budget can be very low, even lower than $100$\,GB. 
Reducing the memory limit increases wall-clock time since  \sysname{} has fewer resources for frontier materialization, but Table~\ref{tab:ratebeer_spilling} shows that the increase in time is reasonable and never excessive.

Overall, the three control mechanisms provide complementary options. Thus, users can select a configuration based on the available memory, disk capacity and bandwidth, and desired execution time. Increasing the number of batches has the advantage of not reducing resources, so it can sometimes reduce memory consumption without increasing execution time, but it does not provide direct control on the amount of memory used, so it does not guarantee that memory will remain within a given budget. On the other hand, reducing the memory limit per worker and relying on the memory management capabilities of the DBMS directly controls memory usage and can ensure that memory remains within a small budget.

%% file: sections/conc.tex
\vspace*{-4pt}
\section{Conclusion}
\label{sec:conc}

We presented \sysname{}, a metapath selection layer for RDL that leverages SQL to identify and prune uninformative metapaths before GNN training. \sysname{} computes lightweight statistics and combines mutual information, sampling cost, and coverage into a scoring function that determines the metapaths to prune. Because the pipeline relies only on database statistics and task labels, it is GNN-agnostic and supports both classification and regression tasks within a single framework. \sysname{} supports temporally scoped prediction under a fixed schema and task definition. Adding or removing a table or foreign-key relationship changes the metapath search space, requiring \sysname{} to be rerun on the updated schema and data snapshot. Given the preprocessing costs reported in Table~3, periodic reruns are practical for the evaluated tasks, and we leave incrementally maintaining sampling rules under schema evolution for future work. Experiments on the RelBench benchmark show that \sysname{} improves both training time and accuracy, especially on RelGT,
demonstrating that informed metapath selection nicely complements strong GNN architectures.